\documentclass[prd,aps,twocolumn,floatfix]{revtex4-2}

\usepackage{graphicx,psfrag,mathrsfs}
\usepackage{slashed}
\usepackage{mathrsfs}
\usepackage{amsmath,amsfonts,amssymb,amsthm}
\usepackage{hyperref}
\hypersetup{breaklinks=true}
\usepackage{url}
\usepackage{comment,cancel}
\usepackage{accents}
\usepackage{ulem}
\usepackage{xcolor,bbold}
\usepackage{orcidlink}
\usepackage{enumitem}
\usepackage{soul}

\makeatletter
\newcommand{\sbullet}{%
  \hbox{\fontfamily{lmr}\fontsize{.4\dimexpr(\f@size pt)}{0}
    \selectfont\textbullet}}
\DeclareRobustCommand{\mathbullet}{\accentset{\sbullet}}
\makeatother

\def\logRu{\log\left(-\frac{R}{\mathfrak{u}}\right)}

\usepackage{amsmath} 
\def\overbigdot#1{\overset{\hbox{\scalebox{0.8}{\tiny$\bullet$}}}{#1}}
\def\mbn{\overbigdot{\nabla}}
\def\Gmmb{\Gamma[\mn,\overbigdot{\nabla}]}
\def\mbbox{\overbigdot{\square}}
\def\Gb{\overbigdot{\Gamma}} 

\def\mn{\mathring{\nabla}}
\def\sD{\slashed{D}}
\def\sD{\slashed{D}} 
\def\sg{\slashed{g}}
\def\psib{\ul{\psi}} 
\def\sigmab{\ul{\sigma}}
\def\cmr{\mathcal{C}^R_-}
\def\cpr{\mathcal{C}^R_+}

\def\p{\partial}

\def\non{\nonumber}
\def\ul{\underline}

\theoremstyle{remark}

\theoremstyle{plain}
\newtheorem{thm}{Theorem}

\allowdisplaybreaks

\begin{document}

\title{An asymptotic systems approach for the good-bad-ugly model with
  application to general relativity}

\author{Miguel Duarte$^{1}$\orcidlink{0000-0003-2223-1304}}
\author{Justin C. Feng$^{2,3}$\orcidlink{0000-0003-2441-5801}}
\author{Edgar Gasper\'in$^1$\orcidlink{0000-0003-1170-5121}}
\author{David Hilditch$^1$\orcidlink{0000-0001-9960-5293}}

\affiliation{$^1$CENTRA, Departamento de F\'isica, Instituto Superior
  T\'ecnico IST, Universidade de Lisboa UL, Avenida Rovisco Pais 1,
  1049 Lisboa, Portugal,
}

\affiliation{$^2$Leung Center for Cosmology and Particle Astrophysics,
  National Taiwan University, No.1 Sec.4, Roosevelt Rd., Taipei 10617,
  Taiwan, Republic of China} \affiliation{$^3$Central European
  Institute for Cosmology and Fundamental Physics, Institute of
  Physics of the Czech Academy of Sciences, Na Slovance 1999/2, 182 21
  Prague 8, Czech Republic}

\begin{abstract}
  We employ an adapted version of H\"ormander's asymptotic systems
  method to show heuristically that the standard
  \textit{good-bad-ugly} model admits formal polyhomogeneous
  asymptotic solutions near null infinity. In a related earlier
  approach, our heuristics were unable to capture potential leading
  order logarithmic terms appearing in the asymptotic solution of the
  good equation (the standard wave equation). Presently, we work with
  an improved method which overcomes this shortcoming, allowing the
  faithful treatment of a larger class of initial data in which such
  logarithmic terms are manifest. We then generalize this method to
  encompass models that include \textit{stratified null forms} as
  sources and whose wave operators are built from an asymptotically
  flat metric. We then apply this result to the Einstein field
  equations in generalized harmonic gauge and compute the leading
  decay in~$R^{-1}$ of the Weyl scalars, where~$R$ is a suitably
  defined radial coordinate. We detect an obstruction to
  \textit{peeling}, a decay statement on the Weyl scalars~$\Psi_n$
  that is ensured by smoothness of null infinity. The leading order
  obstruction appears in~$\Psi_2$ and, in agreement with the
  literature, can only be suppressed by a careful choice of initial
  data.
\end{abstract}

\maketitle

\section{Introduction}

Asymptotically flat spacetimes exhibit common structure at future null
infinity~$\mathscr{I}^+$. In the context of gravitational wave
astronomy, this structure provides a natural framework in which
signals emitted from compact binaries can be understood. Despite
progress~\cite{VanHusHil14,PetGauVan24,MaMoxSch23_a,MaSchMox24}, it is
an open problem to compute such signals at infinity with the methods
of numerical relativity in complete generality. It is therefore of
principle and practical importance to understand the behavior of
asymptotically flat solutions at null infinity.

One of the flagship results of the classical theory of asymptotics is
the peeling theorem which, roughly speaking, states that the curvature
of vacuum asymptotically flat spacetimes decay close to null infinity
as~$\Psi_n = \mathcal{O}(\mathfrak{r}^{5-n})$ where~$n=0,...,4$
and~$\mathfrak{r}$ is an affine parameter along null geodesics ---see
for instance~\cite{Val16}. This theorem makes the underlying
assumption that the spacetime admits a smooth (or at least~$C^4$)
conformal compactification. Although it has been shown from a
rigorous point of view that there exist spacetimes that
peel~\cite{KlaNic03} its clear also that these spacetimes are
non-generic, and there is a large class of non-peeling spacetimes as
discussed in~\cite{ChrMacSin95, Win85, NovGol82, Keh21} and evidenced
by the non-linear stability proofs of~\cite{HinVas20}
and~\cite{ChrKla93}.

In~\cite{DuaFenGas22} a polyhomogeneous peeling result was derived
based on the asymptotic expansions obtained for an equation that we
call the \textit{good-bad-ugly} model in~\cite{GasHil18}. This model
mimics the form of the Einstein field equations (EFE) in harmonic gauge with
the addition of constraints according to the prescription
of~\cite{GasHil18}. This model and extensions of it have proven to be
an ideal arena in which to test regularization strategies for
numerical evolutions that reach~$\mathscr{I}^+$~\cite{GasGauHil19,
  PetGauRaiVanHil23} with the hyperboloidal-dual-foliation approach.
Furthermore, the heuristic asymptotic expansions and formulation of
the Einstein field equations in~\cite{DuaFenGasHil21, DuaFenGasHil22a}
constitute the basic framework for the non-linear numerical evolutions
in spherical symmetry of~\cite{PetGauVan24}. Interestingly,
in~\cite{DuaFenGas22} the peeling violations appear formally at the
same order as those reported in~\cite{GasVal18} obtain using conformal
methods and Friedrich's cylinder at spatial infinity framework. Hence
a natural question that arises is if these logarithmic terms are the
same or not. This question was answered in the negative
in~\cite{DuaFenGasHil23}, where it was shown by taking as base case
the good-bad-ugly model but using conformal methods, that even the
good field which is reported to be log-free in~\cite{DuaFenGasHil21},
contains logarithmic terms.  Hence the question becomes whether the
heuristic asymptotic system method of~\cite{DuaFenGasHil21} is able to
capture the logarithmic terms found with the conformal method in flat
spacetime in~\cite{DuaFenGasHil23} (see also~\cite{Gas24} for a
concise discussion on this issue). The purpose of this paper is to
address this question, albeit heuristically. Although we do not prove
rigorously that these logarithmic terms are the same, we find that
one can indeed devise a heuristic expansion similar to that
of~\cite{DuaFenGas22} that renders logarithmic terms even for the good
field in flat spacetime. The proposed method is moreover flexible
enough to accommodate an analysis of the Einstein field equations that
retrieves asymptotic expansions for the metric components of
asymptotically flat spacetimes. We use these expansions to revisit the
peeling result of~\cite{DuaFenGas22} where, with the more
sophisticated heuristics provided here, we find that there are
logarithmic terms in the gravitational wave degrees of freedom result
in a violation of peeling in~$\Psi_2$. In accordance with the
literature, there appears to be no easy way to avoid this violation
beyond a careful choice of initial data.

The paper is structured as follows. In
section~\ref{section:geometric_setup} we describe the geometric
setup. In section~\ref{section:MK} we present our analysis of the
model in the Minkowski spacetime. In
section~\ref{section:AF_spacetimes} we then extend this treatment to
asymptotically flat spacetimes and consider, in
section~\ref{section:GHG}, the generalized harmonic formulation (GHG)
of general relativity (GR) in this language. In
section~\ref{section:Peeling} we use our heuristic construction to
investigate peeling with the GHG formulation. We conclude in
section~\ref{section:Conclusions}. Geometric units are employed
throughout.

\section{Geometric setup}
\label{section:geometric_setup}

We begin our discussion by introducing relevant notation, the
derivatives we will be working with and the model equations for whose
solutions we want to find the asymptotic behavior near null
infinity. Latin indices will be used as abstract tensor indices,
whereas Greek indices will be used to denote spacetime coordinate
indices. Let~$(\mathcal{M},g_{ab})$ denote a 4-dimensional manifold
equipped with a Lorentzian metric and an associated Levi-Civita
connection~$\nabla$. Let the coordinate
system~$X^{\ul{\alpha}}=(T,X^{\ul{i}})$ be asymptotically Cartesian
on~$\mathcal{M}$ and have an associated flat covariant
derivative~$\mn$ with the defining property
that~$\mn_a\p_{\ul{\alpha}}^b=0$. Let~$\p_{\ul{\alpha}}$
and~$dX^{\ul{\alpha}}$ be the corresponding vector and co-vector
bases. We also define the shell coordinate
system~$X^{\ul{\alpha}'}=(T',X^{\ul{i}'})=(T,R,\theta^A)$, where the
radial coordinate~$R$ is related to~$X^{\ul{i}}$
as~$R^2=(X^{\ul{1}})^2+(X^{\ul{2}})^2+(X^{\ul{3}})^2$.
Let~$\p_{\ul{\alpha}'}$ and~$dX^{\ul{\alpha}'}$ be the corresponding
vector and co-vector bases. Shell coordinates have an associated flat
covariant derivative~$\mbn$, with the defining property
that~$\mbn_b\p_{\ul{\alpha}'}^a=0$. The flat covariant
derivatives~$\mn$ and~$\mbn$ reduce to partial derivatives when
working in their associated coordinates bases. We define the
Christoffel transition tensor between the two derivatives by,
\begin{align}
\Gmmb_{a}{}^{b}{}_{c}v^c = \mn_av^b - \mbn_av^b\,,
\end{align}
with~$v^a$ an arbitrary vector field. In the same way,
\begin{align}
\Gamma[\nabla,\mbn]_{a}{}^{b}{}_{c}v^c = \nabla_av^b - \mbn_av^b\,,
\end{align}
which will be represented with the shorthand~$\Gb_{a}{}^{b}{}_{c} :=
\Gamma[\nabla,\mbn]_{a}{}^{b}{}_{c}$. It follows that,
\begin{align}\label{eq:CartesianAndShellConnectionRelation}
  \Gamma[\nabla,\mn]_b{}^a{}_c =\Gamma[\nabla,\mbn]_b{}^a{}_c -
  \Gamma[\mn,\mbn]_b{}^a{}_c.
\end{align} 
Finally, indices will be raised and lowered with the spacetime
metric~$g_{ab}$ unless otherwise stated.

\subsection{Representation of the metric} 

We define the following outgoing and incoming radial null vector
fields,
\begin{align}\label{eq:psiinshellchart}
  \psi^a&=\p_T^a+\mathcal{C}_+^R\p_R^a\,,\nonumber\\
  \ul{\psi}^a&=\p_T^a+\mathcal{C}_-^R\p_R^a\,,
\end{align}
where~$\mathcal{C}_+^R$ and~$\mathcal{C}_-^R$ are radial coordinate
light-speeds and are determined by the condition that~$\psi^a$
and~$\ul{\psi}^a$ are null with respect to the spacetime
metric. Additionally we define the following covector fields,
\begin{align}\label{eq:xitoeta}
\sigma_a&=e^{-\varphi}\psi_a\,,\quad
\ul{\sigma}_a=e^{-\varphi}\ul{\psi}_a\,,
\end{align}
where~$\varphi$ is fixed by requiring
that~$\sigma_a\p_R^a=-\ul{\sigma}_a\p_R^a=1$. One can
straightforwardly verify the relationships,
\begin{align}\label{eq:xiinshellchart}
  \sigma_a&=-\mathcal{C}_+^R\nabla_aT+\nabla_aR +
  \mathcal{C}^+_A\nabla_a\theta^A\,,\nonumber\\
  \ul{\sigma}_a&=\mathcal{C}_-^R\nabla_aT-\nabla_aR
  + \mathcal{C}^-_A\nabla_a\theta^A\,.
\end{align}
We also have,
\begin{align}
  &\sigma_a\psi^a=\ul{\sigma}_a\ul{\psi}^a=0\,,\non\\ 
  &\sigma_a\ul{\psi}^a=\ul{\sigma}_a\psi^a=-\tau\,,
\end{align}
where~$\tau:=\mathcal{C}_+^R-\mathcal{C}_-^R$. We decompose the
inverse spacetime metric as,
\begin{align}\label{eq:metricrepresentation}
g^{ab}=-2\tau^{-1}e^{-\varphi}\,\psi^{(a}\ul{\psi}^{b)}+\slashed{g}^{ab}
\,,
\end{align}
with the normalization of first term chosen so
that~$\slashed{g}^{ab}\sigma_b=\slashed{g}^{ab}\ul{\sigma}_b=0$. From
this we can also decompose the metric in covariant form as,
\begin{align}
  g_{ab}=-2\tau^{-1}e^{\varphi}\,\sigma_{(a}\ul{\sigma}_{b)}+\slashed{g}_{ab}\,.
\end{align}
It will be convenient to make a conformal rescaling of this induced
metric on the 2-sphere defined by constant values of the
coordinates~$T$ and~$R$. We define,
\begin{align}\label{eq:qconformaltransformation}
  q_{ab}= e^{-\epsilon}R^{-2}\slashed{g}_{ab}, \qquad (q^{-1})^{ab} =
  e^{\epsilon}R^{2}\slashed{g}^{ab}
\end{align}
with,
\begin{align}
\epsilon = (\ln|\slashed{g}|-\ln|\mathbullet{\slashed{g}}| )/2
\end{align}
where~$ |\mathbullet{\slashed{g}}|$ denotes the determinant of the
metric of~$\mathbb{S}^2$ of radius~$R$ embedded in Minkowski spacetime
in shell coordinates. The tensor~$(q^{-1})^{AB}$ can be decomposed as
a function of the two degrees of freedom of gravitational waves~$h_+$
and~$h_\times$ as,
\begin{align}
(q^{-1})^{AB} = \begin{bmatrix}
e^{-h_+}\cosh h_\times & \frac{\sinh h_\times}{\sin \theta} \\
\frac{\sinh h_\times}{\sin \theta}
& \frac{e^{h_+}\cosh h_\times}{\sin \theta ^2}
\end{bmatrix}\,.
\end{align}
We can now write the metric as a function of the following ten
independent variables,
\begin{align}\label{eq:BasicMetricVariables}
 \varphi\,, \quad\mathcal{C}_\pm^R\,, \quad \mathcal{C}^\pm_A\,, \quad
 \epsilon\,, \quad h_+\,, \quad h_\times\,.
\end{align}
Tensors~$T'_{cd}$ projected with~$\slashed{g}{}_a{}^b$
are~$\slashed{T}_{ab} \equiv
\slashed{g}{}_a{}^c\slashed{g}{}_{b}{}^dT'_{cd}$. The projected
covariant derivative is denoted~$\sD$, and is given explicitly below
for vectors satisfying~$v^a=\slashed{g}{}_{b}{}^{a}v^b$,
\begin{align}
\sD_b v^a := \slashed{g}{}_{c}{}^{a}\slashed{g}{}_{b}{}^{d}\nabla_d v^c\,.
\end{align}
For a more detailed explanation of this particular way to represent
the metric we refer the reader to~\cite{DuaFenGasHil22a}.

\subsection{3+1 decomposition}

Let~$\mathcal{S}$ be a hypersurface defined by the
condition~$T=0$. Throughout this work, this is the slice on which we
will be assigning initial data. The spacetime metric~$g_{ab}$ induces
a metric on~$\mathcal{S}$, $\gamma_{ab}$ with an associated covariant
derivative~$D_a$. Later on it will be useful to have~$\gamma_{ij}$
written explicitly in our variables, as well as the corresponding
lapse and shift, $\alpha$ and~$\beta_i$, respectively. We have,
\begin{align}\label{inducedMetric}
\gamma_{ij} = \begin{bmatrix}
-\frac{2e^\varphi}{\tau} & \frac{e^\varphi(\mathcal{C}^-_A-\mathcal{C}^+_A)}{\tau} \\ 
\frac{e^\varphi(\mathcal{C}^-_B-\mathcal{C}^+_B)}{\tau}& q_{AB}
\end{bmatrix}\,,
\end{align}
with,
\begin{align}
&q_{\theta^1\theta^1} = e^{\epsilon+h_+}\cosh(h_\times)R^2 
 +\frac{2e^\varphi\mathcal{C}^-_{\theta^1}\mathcal{C}^+_{\theta^1}}{\tau}\,,\non\\	
&q_{\theta^1\theta^2} = -e^\epsilon\sinh h_\times R^2\sin\theta^1 
 +\frac{e^\varphi}{\tau}\left(\mathcal{C}^-_{\theta^1}\mathcal{C}^+_{\theta^2}
 +\mathcal{C}^-_{\theta^2}\mathcal{C}^+_{\theta^1}\right)\,,\non	\\
&q_{\theta^2\theta^2} =  e^{\epsilon-h_+}\cosh h_\times\sin^2 \theta^1
 +\frac{2e^\varphi\mathcal{C}^-_{\theta^2}\mathcal{C}^+_{\theta^2}}{\tau}\,.
\end{align}
The lapse and shift are given by,
\begin{align}\label{LapseShift}
  &\alpha=-\frac{1}{\sqrt{-g^{TT}}}\,,\non\\
  &\beta_i =-\frac{e^\varphi}{\tau}
\begin{bmatrix}
 \cpr+\cmr \,\,\,,& \cmr\mathcal{C}^+_A-\cpr\mathcal{C}^+_A
\end{bmatrix}\,,
\end{align}
where the first expression was not written explicitly because it is
rather long and not particularly enlightening for our purposes.

\subsection{Wave operators}

Our model consists of a system of inhomogeneous wave equations, so we
start by defining the wave operators that will be used. We
define~$\square$ as,
\begin{align}
  \square \phi = g^{ab}\nabla_a\nabla_b\phi\,,
\end{align}
and we define~$\mathring{\square}$ and $\mbbox$ analogously,
associated with~$\mn$ and~$\mbn$, respectively. Note that when
$\mathring{\square}$ acts on a scalar function, one can
straightforwardly change to~$\mbbox$ through,
\begin{align}
  \mathring{\square} \phi = \mbbox \phi -
  g^{bc}\Gamma[\mathring{\nabla},
    \mbn]_b{}^{a}{}_c\nabla_a\phi\,.
\end{align}
Moreover, in~\cite{DuaFenGasHil21} it was shown
that~$\mathring{\square}$ can be expanded as,
\begin{align}\label{lhs2}
  \mathcal{C}_+^R\mathring{\square} \phi = &-
  2e^{-\varphi}\nabla_{\psi}\nabla_{T} \phi
  +\nabla_T\phi(\slashed{g}^{ab}\Gmmb_{a}{}^{\sigma}{}_{b} + X_T)\non\\
  &+\nabla_{\psi}\phi X_\psi
  -\frac{2e^{-\varphi}\mathcal{C}_-^R}{\tau}\nabla_{\psi}^2\phi
  + \mathcal{C}_+^R\cancel{\Delta}\phi\,,
\end{align}
where~$X_T$ and~$X_\psi$ are,
\begin{align}
  \tau X_T:= &\mathcal{C}_A\sD^A\mathcal{C}_+^R
  - \tau\sD^A\mathcal{C}_A^+
  - \frac{2e^{-\varphi}\mathcal{C}_-^R}{\mathcal{C}_+^R}
  \nabla_\psi\mathcal{C}_+^R\,,\\\non
  \tau X_\psi:=& \frac{\mathcal{C}_A}{\tau}\left(\mathcal{C}_-^R
  \sD^A\mathcal{C}_+^R-\mathcal{C}_+^R
  \sD^A\mathcal{C}_-^R\right)
  - \mathcal{C}_-^R\sD^A\mathcal{C}_A^+\non\\
  &- \mathcal{C}_+^R\sD^A\mathcal{C}_A^-
  + \mathcal{C}_-^R\slashed{g}^{ab}\Gmmb_{a}{}^{\sigma}{}_{b}\non\\
  &+ \mathcal{C}_+^R\slashed{g}^{ab}\Gmmb_{a}{}^{\ul{\sigma}}{}_{b}
  + \frac{2e^{-\varphi}\mathcal{C}_-^R}{\mathcal{C}_+^R}
  \nabla_\psi\mathcal{C}_+^R \,,\non
\end{align}
and~$\mathcal{C}_A:=\mathcal{C}_A^++\mathcal{C}_A^-$. Here, the
vector~$T^a$ is simply shorthand for~$\p_T{}^a$.

\subsection{Model equations} 

\textit{Stratified null forms} (SNFs) are expressions involving
products of terms containing at most first derivatives of evolved
fields and decaying close to null infinity faster
than~$R^{-2}$. Logically, this decay cannot be known before
integrating the equations, so this is based on the standard
assumptions that good derivatives (tangent to the outgoing null cone)
of solutions to wave equations decay at least one order faster than
the solutions themselves, whereas bad derivatives decay only as fast
as the solutions~\cite{DuaFenGasHil21}. Throughout this work we will
use a calligraphic~$\mathcal{N}_\phi$ to denote stratified null forms,
where~$\phi$ is the field whose evolution equation
contains~$\mathcal{N}_\phi$.  In this paper we are concerned with a
generalized version of the so-called \textit{good-bad-ugly} system, as
defined in~\cite{DuaFenGasHil21},
\begin{align}
  & \mathring{\square} g = \mathcal{N}_g\,,\non\\ &\mathring{\square}
  b = (\nabla_T g)^2 + \mathcal{N}_b\,,\non\\ &\mathring{\square} u =
  \tfrac{2}{R}\nabla_T u + \mathcal{N}_u\,,
  \label{gbu}
\end{align}
where~$g$, $b$ and~$u$ stand for~{\it good, bad} and {\it ugly}
fields, respectively.

\subsection{Assumptions}
\label{subsection:Assumption}

We define a set of functions from the metric components that we denote
by variations of~$\gamma$,
\begin{align}
  &\varphi = \gamma_1\,,\quad
    \mathcal{C}_\pm^R = \pm1
    + \gamma^\pm_2\,,\quad
    \mathcal{C}_A^\pm = R\gamma_3^\pm\,, \non \\
  &\epsilon = \gamma_4\,,\quad
    h_+ = \gamma_5\,,\quad
    h_\times = \gamma_6\,.\label{asympflatness}
\end{align}
At null infinity, we assume that the $\gamma$ functions, as well
as~$g$, $b$ and~$u$ satisfy,
\begin{align}\label{decayScri}
	\gamma=o^+(1),\, g=o^+(1),\, b=o^+(1), \, u=o^+(1)\,,
\end{align}
where the notation~$f=o^+(h)$ means,
\begin{align}\label{deflittleo-Est}
  \exists \epsilon>0 : \lim_{R\rightarrow\infty}
  \frac{f}{hR^{-\epsilon}}=0\,,
\end{align}
which can be informally stated as \textit{$f$ falls-off faster
  than~$h^{1+\epsilon}$ as $R$ goes to infinity for any $\epsilon>0
  $}, which is a faster decay than~$f=o(h)$. Additionally, we assume
that the $\gamma$ functions depend analytically on $g$, $b$ and
$u$. Throughout this work we will focus mainly on a specific class of
initial data given by,
\begin{align}\label{ID}
  \phi\rvert_{\mathcal{S}} =
  \sum_{n=1}^{\infty}\frac{\bar{M}_{\phi,n}}{R^n}\,,\quad
  \nabla_T\phi\rvert_{\mathcal{S}}=\sum_{n=1}^{\infty}\frac{M_{\phi,n}}{R^{n+1}}\,,
\end{align}
where both~$M_{\phi,n}$ and~$\bar{M}_{\phi,n}$ are functions of only
the angles in the shell coordinate system. The word \textit{order}
will be used to refer to the powers of~$R^{-1}$. Finally, we assume
that derivatives along~$\psib^a$ of~$\gamma_2^+$ decay at least one
order faster than~$\gamma_2^+$ itself near null infinity. This will be
important when we treat asymptotically flat metrics.

\section{Minkowski spacetime}
\label{section:MK}

In this section we want to study the standard version of the
good-bad-ugly model. This means the case where the
background~$(\mathcal{M},g_{ab})$ is the Minkowski spacetime and there
are no SNFs. Namely,
\begin{align}
  & \mathring{\square} g = 0\,,\non\\
  &\mathring{\square} b = (\nabla_T g)^2 \,,\non\\
  &\mathring{\square} u = \tfrac{2}{R}\nabla_T u \,.
  \label{gbuFlat}
\end{align}
where the ten metric functions take the following values,
\begin{align}
	&\mathcal{C}_\pm^R = \pm1\,,\non\\
	&\varphi=\mathcal{C}_A^\pm = 0\,, \non\\
	&\slashed{g}^{ab} = \slashed{\eta}^{ab}\,,\label{flatmetricfuncs}
\end{align}
and the only unknowns are the evolved fields. Additionally, $\psi^a$
and~$\ul{\psi}^a$ reduce to,
\begin{align}
  &\psi^a=\p_T^a+\p_R^a\,,\non\\
  &\ul{\psi}^a=\p_T^a-\p_R^a\,.\label{eq:xiinshellchartflat}
\end{align}
In line with the method laid out in~\cite{DuaFenGasHil21}, the
calculations are made easier if we write the incoming null vector
field~$\psib^a$ as a function of the outgoing vector field~$\psi^a$
and the partial derivative with respect to the time coordinate~$T$,
\begin{align}
  \ul{\psi}^a = 2\p_T^a - \psi^a\,.
\end{align}
Throughout this section it is convenient to use the
coordinates~$(\mathfrak{u},R,\theta^A)$ instead of~$(T,R,\theta^A)$,
where $\mathfrak{u}=T-R$. For this reason, the
derivative~$\nabla_\psi$ should be understood as $\p_R$ and $\nabla_T$
should be understood as~$\p_\mathfrak{u}$. The
coordinate~$\mathfrak{u}$ will be defined in a slightly more
complicated way once we have to deal with curved spacetimes, but for
now this will suffice.

\subsection{Good field}

\begin{figure}[t!]
\includegraphics[width=0.45\textwidth]{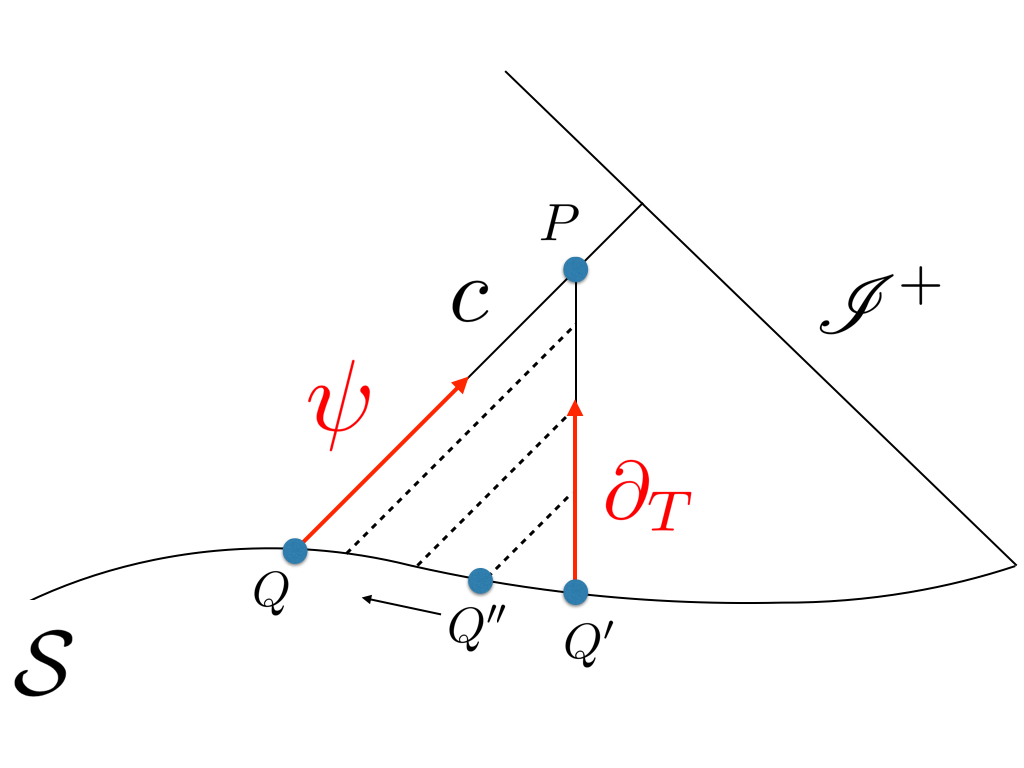}
\caption{A schematic of our geometric setup. The method proceeds first
  by integrating out along~$c$, an integral curve of the outgoing
  null-vector~$\psi^a$ and then up along integral curves of~$\p_T^a$.
  \label{fig:Int_Fig}}
\end{figure}

Rescaling the good field by one power of~$R$ we get,
\begin{align}
  G=gR\,,
\end{align}
and plugging that into the standard flat space wave equation we find
that,
\begin{align}
	\nabla_\psi\nabla_T G=\frac{1}{2}(\nabla_\psi^2+\cancel{\Delta})G\,,
\end{align}
where $\cancel{\Delta}\phi := \mathring{\sD}^a\mathring{\sD}_a\phi $.
We rewrite the field~$G$ as a sum of fields,
\begin{align}
  G=\sum_{n=1}^{\infty}G_n\,,
\end{align}
where the functions~$G_n$ are allowed to depend upon all
coordinates. We get,
\begin{align}\label{good-inf-sum}
  \sum_{n=1}^{\infty}\nabla_\psi\nabla_T G_n
  =\frac{1}{2}\sum_{n=1}^{\infty}(\nabla_\psi^2+\cancel{\Delta})G_n\,.
\end{align}
Note that we get a solution to this equation if we require that each
term on one of the sides equals another on the other side. Naturally,
there is an infinite number of ways to do that, but we can draw some
intuition from H\"ormander's asymptotic systems method in order to
make an educated guess. Considering that derivatives along~$\psi^a$
and angular derivatives are \textit{good derivatives}, meaning that
they increase the decay of their argument by~$R^{-1}$, and that time
derivatives are bad, meaning that they do not increase the decay, we
can easily see that each term on the left-hand side has one of each
and that each term on the right-hand side has two good
derivatives. This suggests that the orders of the terms on the
right-hand side are staggered by 1. With this in mind, we choose to
equate terms in the following way,
\begin{align}\label{good-collected}
  \nabla_\psi\nabla_T
  G_n=\frac{1}{2}(\nabla_\psi^2+\cancel{\Delta})G_{n-1}\,.
\end{align}
Notice that even after finding solutions to the infinite number of
resulting equations, there is no guarantee that the series converges
and, in contrast with the conformal method of \cite{GasMagMen24,
  DuaFenGasHil21}, a partial sum (truncated series) is not an exact
solution. This is the main shortcoming of this heuristic method.
However, as we will discuss later, it is relatively straightforward to
generalize this method for asymptotically flat backgrounds. Our goal
here is to find the high order functional form of the field~$g$ using
induction, so let us turn our attention to the case~$n=1$. We get,
\begin{align}\label{G1}
  \nabla_\psi\nabla_T G_1=0\,.
\end{align}
In order to find the value of the solution~$G_1$ at a generic
point~$P$ in the spacetime, we want to integrate equation~\eqref{G1}
along integral curves of~$\psi^a$ and then~$\p_T{}^a$ as is shown on
Fig.~\ref{fig:Int_Fig}.

Let~$c$ be the integral curve of the vector field~$\psi^a$ that passes
through the point~$Q \in \mathcal{S}$ and integrate
equation~\eqref{G1} along that curve from $Q$ to $P$ to get,
\begin{align}\label{goodfirstint1}
  \nabla_T G_1 = (\nabla_T G_1)|_Q\,,
\end{align}
where it is understood that the left-hand side of equation
\eqref{goodfirstint1} is evaluated at
some~$P=(\mathfrak{u}^*,R^*, \theta^{A*})$. Then, since integral
curves of $\psi^a$ are characterized by constant values
of~$\mathfrak{u}$ and $\mathcal{S}$ is defined by the condition $T=0$,
we must have~$Q=(\mathfrak{u}^*,-\mathfrak{u}^*,\theta^{A*})$. On the
other hand, integral curves of~$\p_T{}^a$ have the property that~$R$
is held constant, so~$Q'=(-R^*,R^*, \theta^{A*})$. Therefore,
evaluating a function at the point~$Q$ amounts to requiring
that~$R=-\mathfrak{u}^*$. Since the curve $c$ and the point~$Q$ are
fixed but arbitrary, equation~\eqref{goodfirstint1} is valid for any
integral curve~$c$ of~$\psi^a$ intersecting~$\mathcal{S}$. Other
integral curves of~$\psi^a$ are depicted in Fig.~\ref{fig:Int_Fig}. as
dashed lines. The integration along~$\p_T{}^a$ will then run through
these curves, effectively dragging~$Q''$ from~$Q'$ to~$Q$. The lower
integration limit of the first integral is then dependent on the
integration variable of the second integral and determined by the
shape of the integral curves. The latter will become more complex as
we allow the spacetime to be curved, but we leave that discussion for
the next section. We can write~$\mathfrak{u}^*=\mathfrak{u}$
and~$R^*=R$. Also,
\begin{align}\label{goodfirstint2}
  \nabla_T G_1 = (\nabla_T G_1)|_{R\rightarrow -\mathfrak{u}}\,.
\end{align}
The exact same method will be used in every integration throughout the
paper. Integrating along~$\p_T{}^a$ we get,
\begin{align}\label{goodsecondint}
  G_1 = \int^{\mathfrak{u}}_{-R}(\nabla_T G_1)|_{R\rightarrow
  -\mathfrak{u}}d\mathfrak{u}'+G_1\rvert_{Q'}\,,
\end{align}
or rather,
\begin{align}\label{goodsecondint2}
  G_1 = \int^{\mathfrak{u}}_{-R}(\nabla_T G_1)|_{R\rightarrow -\mathfrak{u}}
  d\mathfrak{u}'+G_1\rvert_{\mathfrak{u}\rightarrow -R}\,.
\end{align}
We now have an expression for the field~$G_1$ as a function of initial
data, but it is important to note that~$G_1$, much like all~$G_n$,
have no physical meaning by themselves. They are auxiliary functions
which, added together and properly rescaled, make up our physical
field~$g$. So there is no unique or obvious way to split our initial
data for~$g$, equation~\eqref{ID}, among the different~$G_n$. It is a
choice that we have to make at this point. Rescaling a generic
field~$\phi$ the same way we rescaled the field~$g$, namely,
\begin{align}
  \Phi=\phi R\,,
\end{align}
let us choose the following,
\begin{align}\label{IDsplit}
  \Phi_n\rvert_{\mathcal{S}} = \frac{\bar{M}_{\phi,n}}{R^{n-1}}\,,\quad
  (\nabla_T\Phi_n)\rvert_{\mathcal{S}}=\frac{M_{\phi,n}}{R^{n}}\,,
\end{align}
where~$\bar{M}_{\phi,n}$ and~$M_{\phi,n}$ are functions of~$\theta^A$
only.  Plugging~\eqref{IDsplit} in~\eqref{goodsecondint2} we find,
\begin{align}
  \label{G1Final}
  G_1 = \bar{M}_{g,1}+M_{g,1}\logRu\,.
\end{align}
Having integrated the expression for $G_1$ we are now ready to
approach the general case $G_n$. Integrating~\eqref{good-collected}
along integral curves of $\psi^a$ in exactly the same way we get,
\begin{align}\label{GnfirstInt}
  \nabla_T G_n&=\frac{1}{2}\int_{-\mathfrak{u}'}^R
  (\nabla_\psi^2+\cancel{\Delta})G_{n-1}dR'
  + \left(\nabla_T G_n\right)\rvert_{R\rightarrow -\mathfrak{u}}\,,
\end{align}
and integrating along~$\p_T{}^a$,
\begin{align}\label{GnsecondInt}
G_n=&\frac{1}{2}\int_{-R}^\mathfrak{u}\int_{-\mathfrak{u}'}^R
      (\nabla_\psi^2+\cancel{\Delta})G_{n-1}dR'd\mathfrak{u}'\\
    &+ \int_{-R}^\mathfrak{u}\left(\nabla_T G_n\right)
      \rvert_{R\rightarrow -\mathfrak{u}'}
      d\mathfrak{u}'+\left(G_n\right)\rvert_{\mathfrak{u}\rightarrow -R}\,.\non
\end{align}
Notice the last expression provides a recursive way to compute~$G_n$
in terms of~$G_{n-1}$ where~$R'$ represents the parameter on each of
the curves of a 1-parameter family of curves labeled (parameterized)
by~$\mathfrak{u}'$. Let us focus first on the terms we can get from
initial data, i.e. the second line of~\eqref{GnsecondInt}. The first
term gives,
\begin{align}\label{GnIDTerms1}
  \int_{-R}^\mathfrak{u}\left(\nabla_T G_n\right)
  &\rvert_{R\rightarrow
  -\mathfrak{u}'}d\mathfrak{u}'\\ 
  &=\frac{M_{g,n}}{n-1}\left[\frac{1}{(-\mathfrak{u})^{n-1}}
    -\frac{1}{R^{n-1}}\right],\non
\end{align}
for all $n>1$, whereas the second gives,
\begin{align}\label{GnIDTerms2}
	G_n\rvert_{\mathfrak{u}\rightarrow -R} = \frac{\bar{M}_{g,n}}{R^{n-1}}\,.
\end{align}
The only thing left to do is to deal with the first term on the
right-hand side of~\eqref{GnsecondInt}. This term is significantly
more intricate because it is the one that makes~\eqref{GnsecondInt} a
recurrence relation. We must then find a functional form for~$G_{n}$
which is kept the same for~$G_{n+1}$ and, at the same time,
includes~$G_1$. Computing the first few~$G_n$ in computer algebra we
can get a sufficiently detailed understanding of the type of terms
that emerge every time we increase~$n$ by~$1$. We assume that~$G_{n}$
behaves as,
\begin{align}\label{GnHyp}
  G_n=&\sum_{j=0}^{n-1}\sum_{q=j-n}^{j}A_{j,q,0}(\theta^A)
        \frac{\mathfrak{u}^q}{R^j}\\ 
      &+\sum_{j=0}^{n-1}\sum_{q=0}^{j}A_{j,q,1}(\theta^A)
        \frac{\mathfrak{u}^q}{R^j}\logRu\,.\non
\end{align}
Then we plug~\eqref{GnHyp} into the first term on the right-hand side
of~\eqref{GnsecondInt} and check that it takes the form,
\begin{align}
  G_{n+1}=&\sum_{j=0}^{n}\sum_{q=j-n-1}^{j}\bar{A}_{j,q,0}(\theta^A)
            \frac{\mathfrak{u}^q}{R^j}\\ 
          &+\sum_{j=0}^{n}\sum_{q=0}^{j}\bar{A}_{j,q,1}(\theta^A)
            \frac{\mathfrak{u}^q}{R^j}\logRu\,,\non
\end{align}
for other angular functions~$\bar{A}_{j,q,i}(\theta^A)$, as
desired. We omit the details of the integration, since the expressions
are long and the calculation is relatively straightforward. Next, we
can easily check that, for certain choices of~$A_{j,q,i}(\theta^A)$,
the expression on the right-hand side of~\eqref{GnHyp} turns into the
terms coming from assumptions on initial data~\eqref{GnIDTerms1}
and~\eqref{GnIDTerms2}. Finally, it is also trivial to verify
that~\eqref{G1Final} is included in the right-hand side
of~\eqref{GnHyp}. This concludes the proof. Notice that regardless of
the value that~$n$ takes, $j$ runs from~$0$ to~$n-1$, which means that
in general there will be leading order contributions at every
step. This is counter-intuitive, since one would expect this type of
construction would identify the parameter~$n$ with the order at
which~$G_n$ contributes, or rather that~$j=n-1$. However, if one takes
a closer look at the integration limits in~\eqref{GnsecondInt}, it
becomes obvious that each~$G_n$ must come, in general, with
contributions at every order, since terms that only
have~$\mathfrak{u}$ and no~$R$ are necessarily present. This is
unfortunate for the purpose of this work, but it only serves to show
that the type of stratification we can more naturally expect involves
the behavior in~$\mathfrak{u}$ as well as in~$R$. To make this point
clearer we consider a function $F$ that is of the form,
\begin{align}
  F=A(\theta^A)\frac{\mathfrak{u}^q}{R^j}\,,
\end{align} 
where~$q$ and~$j$ are such that $j-q=m-1$, and we apply to it the same
derivatives and integrals as we do in the first term on the right-hand
side of~\eqref{GnsecondInt},
\begin{align}\label{Powers}
  \int_{-R}^\mathfrak{u}\int_{-\mathfrak{u}'}^R(\nabla_\psi^2+
  &\cancel{\Delta})A\frac{\mathfrak{u}^q}{R^j}dR'd\mathfrak{u}'\\
  &=\bar{A}\frac{\mathfrak{u}^{q+1}}{R^{j+1}}+\tilde{A}\mathfrak{u}^{1-m}
    +A'\frac{1}{R^{m-1}}\non\,,
\end{align}
where~$\bar{A}$, $\tilde{A}$ and~$A'$ are functions of the
angles. Looking closely at~\eqref{Powers}, we see that all 3 terms on
the right-hand side share the property that~$j-q=m-1$. So what the
structure of this method preserves is not a relationship between~$n$
and the power of~$R$, $j$, but a relationship between the latter and
the power of~$\mathfrak{u}$, $q$. Nevertheless, since we have already
shown~\eqref{GnHyp}, we can reorganize the terms in order to identify
the order in~$R$ with the parameter~$n$ by simply
redefining~$G_n$. Since we know that, in general, each order has at
most one power of~$\log(-R/\mathfrak{u})$, our good field takes the
schematic form,
\begin{align}
  g=\sum_{n=1}^\infty\frac{\mathcal{G}_{n,0}(\psi^*)
  +\mathcal{G}_{n,1}(\psi^*)\logRu}{R^n}\,,
\end{align}
where~$\mathcal{G}_{n,0}(\psi^*)$ and~$\mathcal{G}_{n,1}(\psi^*)$ are
functions that do not vary along integral curves of the outgoing null
vector field. In Minkowski spacetime this is just another way of
saying that the function is not allowed to depend on~$R$, but we
introduce this notation because that is not necessarily the case in
curved spacetimes.

\subsection{On the generality of the first order log}

Our choice of initial data~\eqref{ID} is quite strict because we
wanted to be able to do the necessary integrations up to any
order. However, making a choice this specific might raise doubts about
how general the logarithm that appears at leading order in solutions
to the standard wave equation really is. Is this something that
appears as a result of very particular choices or are we likely to
have to deal with such logs for a very large class of initial data? In
this subsection we want to argue that the latter is true. Let us then
choose the following,
\begin{align}\label{IDgeneral}
  \nabla_TG_1\rvert_{\mathcal{S}}=\frac{M_{g,1}(\mathfrak{u},R,\theta^A)}{R}\,,
\end{align}
where~$M_{g,1}$ is now allowed to depend upon all four coordinates and
let us also not specify anything else regarding how our solution
behaves on the initial slice. As in the last subsection, the
equation~$G_1$ must satisfy is,
\begin{align}
  \nabla_\psi\nabla_T G_1=0\,,
\end{align}
which gives,
\begin{align}\label{G1integralsgeneralID}
  G_1 = \int^{\mathfrak{u}}_{-R}(\nabla_T G_1)|_{R\rightarrow -\mathfrak{u}}
  d\mathfrak{u}'+G_1\rvert_{\mathfrak{u}\rightarrow -R}\,.
\end{align}
Replacing~\eqref{IDgeneral} in~\eqref{G1integralsgeneralID} we get,
\begin{align}\label{G1integralsgeneralID2}
  G_1 = \int^{\mathfrak{u}}_{-R}\frac{M_{g,1}(\mathfrak{u}',-\mathfrak{u}',\theta^A)}
  {\mathfrak{u}'}d\mathfrak{u}'+G_1\rvert_{\mathfrak{u}\rightarrow -R}\,.
\end{align}
Looking at the first term on the right-hand side
of~\eqref{G1integralsgeneralID2}, we see that different choices of
functions~$M_{g,1}$ give a logarithm after the integration. One
example besides a purely angular function is any function with
symmetric powers of~$R$ and~$\mathfrak{u}$. Note that we also have not
specified any information on the choice of how~$G_1$ behaves on the
initial slice and we have not mentioned the behavior of
other~$\nabla_T G_n$ and~$G_n$ either. This suggests that this leading
order logarithm in the solution to the wave equation is not only not
specific to our choice of initial data, but indeed quite a general
feature.

It is interesting to pause here and analyze the expressions we got in
this subsection. It seems that the functional form of the solution to
the wave equation has a leading order logarithm, see
equation~\eqref{G1Final}, with the class of initial data that we
chose~\eqref{ID}. This is in agreement with the leading order behavior
of the solutions found in~\cite{DuaFenGasHil23}. As the methods are
substantially different, one cannot, without further research, assert
with confidence that they are the same logs. However, we can check
whether the requirement found in~\cite{DuaFenGasHil23} (at leading
order) for this log to vanish makes the one in~\eqref{G1Final} vanish
as well.  The class of initial data used in~\cite{DuaFenGasHil23} is a
subclass of ours~\eqref{ID} in the sense that each~$M_{\phi,n}$
and~$\bar{M}_{\phi,n}$ can only contain a finite number of spherical
harmonics~$Y_{\ell m}$ with~$0 \leq \ell \leq n-1$. There, the authors
find that requiring~$M_{g,1}=0$ is a necessary and sufficient
condition for the leading log to vanish and it is easy to see that the
same statement can be said of our expression
for~$G_1$~\eqref{G1Final}. More recently, in~\cite{GasMagMen24} the
restriction on initial data made in~\cite{DuaFenGasHil23} is lifted to
allow for~$M_{\phi,n}$ and~$\bar{M}_{\phi,n}$ to be expressed as
infinite sums of spherical harmonics~$l=0,...,\infty$ and the solution
is computed using conformal methods.  Although a more detailed
comparison between the asymptotic system approach put forward in this
article and the conformal method of~\cite{GasMagMen24}
and~\cite{DuaFenGasHil23} is necessary to assess whether the
logarithmic terms are the same or not, our expectation is that they
are.  This expectation comes from the fact that we have observed
through computer algebra that at least up to~$n=10$ and~$l=5$, the
necessary and sufficient condition to eliminate logarithmic terms
coincides with the one stated in \textit{Proposition 1}
of~\cite{GasMagMen24}.

\subsection{Bad field}

We now turn our attention to the bad equation. Rescaling the field in
exactly the same way as we rescaled the good,
\begin{align}
  B=bR\,,
\end{align}
and substituting that into the bad equation in flat space we get,
\begin{align}
  \nabla_\psi\nabla_T
  B=\frac{1}{2}(\nabla_\psi^2+\cancel{\Delta})B
  -\frac{R}{2}(\nabla_T g)^2\,.
\end{align}
Rewriting the rescaled bad field as an infinite sum of fields $B_n$,
\begin{align}
  B=\sum_{n=1}^\infty B_n\,,
\end{align}
where~$B_n$ is allowed to depend on every coordinate, we find,
\begin{align}\label{bad-inf-sum}
  \sum_{n=1}^{\infty}\nabla_\psi\nabla_T
  B_n=\frac{1}{2}\sum_{n=1}^{\infty}(\nabla_\psi^2
  +\cancel{\Delta})B_n-\frac{R}{2}(\nabla_T g)\,.
\end{align}
We can now equate terms on either side using the same reasoning as in
the previous subsection, but we have not yet made a choice on how to
do that with the second term on the right-hand side. We choose to
collect terms order by order, meaning,
\begin{align}
  \label{bad-collected}
  \nabla_\psi\nabla_T
  B_n=\frac{1}{2}(\nabla_\psi^2+\cancel{\Delta})B_{n-1}-\frac{1}{2R^n}C_n\,,
\end{align}
where $C_n$ is defined as,
\begin{align}
  \label{Cn}
  &C_n = \sum_{i,j=1}^{i+j=n+1}\nabla_T G_i\nabla_T G_j\,.
\end{align}
Here,~$\sum_{i,j=1}^{i+j=n+1}$ is meant as the \textit{sum over terms
  with any combination of~$i$ and~$j$ as long as~$i,j\geq 1$
  with~$i+j=n+1$}. Let us deal first with the case
$n=1$. From~\eqref{bad-collected} we get,
\begin{align}
  \nabla_\psi\nabla_T B_1=\frac{1}{2R}(\nabla_TG_1)^2\,.
\end{align}
Integrating along integral curves of~$\psi^a$ and then
along~$\p_T{}^a$ in the exact same way as for the good field we find,
\begin{align}
  B_1=&\frac{1}{2}\int_{-R}^\mathfrak{u}\int_{-\mathfrak{u}'}^R
        \frac{1}{R}(\nabla_TG_1)^2 dR'd\mathfrak{u}'\\ 
      &+\int^{\mathfrak{u}}_{-R}(\nabla_TB_1)|_{R\rightarrow -\mathfrak{u}}
        d\mathfrak{u}'+B_1\rvert_{\mathfrak{u}\rightarrow -R}\,.\non
\end{align}
Let us choose the same distribution of initial data as for the good
field, namely~\eqref{IDsplit}. The terms that come from initial data,
the second and third on the right-hand side, have the same form as the
ones computed in the previous section, whereas the first term on the
right-hand side can be calculated by substituting~\eqref{G1Final}
directly. We get,
\begin{align}\label{B1Final}
	B_1=&\frac{M_{g,1}^2}{2}
              \left(\frac{1}{\mathfrak{u}}-\frac{1}{R}\right)
              -\frac{M_{g,1}^2}{2\mathfrak{u}}\logRu\non\\ 
            &+\bar{M}_{b,1}+M_{b,1}\logRu\,.
\end{align}
It is worth analyzing this expression in order to understand how the
logarithmic terms are generated. The form that~$B_1$ takes has logs
from two different origins. The one in the second term comes from the
term that contains a time derivative of~$G_1$, but it is not
`inherited' directly from~$G_1$ in the sense that the time derivative
makes that log disappear and it only appears again upon integrating
in~$R$. The log in the last term is analogous to the one present
in~\eqref{G1Final}. Let us now integrate the
expression~\eqref{bad-collected},
\begin{align}\label{BnsecondInt}
  B_n=&\frac{1}{2}\int_{-R}^\mathfrak{u}\int_{-\mathfrak{u}'}^R
        (\nabla_\psi^2+\cancel{\Delta})B_{n-1}dR'd\mathfrak{u}'\non\\ 
      &-\frac{1}{2}\int_{-R}^\mathfrak{u}\int_{-\mathfrak{u}'}^R
        \frac{1}{R^n}C_ndR'd\mathfrak{u}'\\ 
      &+\int_{-R}^\mathfrak{u}\left(\nabla_T B_n\right)
        \rvert_{R\rightarrow-\mathfrak{u}'}
        d\mathfrak{u}'+B_n\rvert_{\mathfrak{u}\rightarrow -R}\,,\non
\end{align}
and begin by computing the last line because it is analogous to what
we have already found in the last subsection,
\begin{align}\label{BnIDTerms1}
  \int_{-R}^\mathfrak{u}\left(\nabla_T B_n\right)
  &\rvert_{R\rightarrow
    -\mathfrak{u}'}d\mathfrak{u}'=\\ 
  &=\frac{M_{b,n}}{n-1}\left[\frac{1}
    {(-\mathfrak{u})^{n-1}}-\frac{1}{R^{n-1}}\right],\non
\end{align}
for all~$n>1$, and,
\begin{align}\label{BnIDTerms2}
  B_n\rvert_{\mathfrak{u}\rightarrow -R} = \frac{\bar{M}_{b,n}}{R^{n-1}}\,.
\end{align}
For the remaining two terms in~\eqref{BnsecondInt} we must find a
suitable induction hypothesis. We want to use something similar to the
hypothesis shown for the good equation~\eqref{GnHyp}, but with one
important change, namely that logs may now appear with powers higher
than one. This is not obvious from the $n=1$ case~\eqref{B1Final},
where logs only appear linearly, but it becomes clear once we notice
that some of the logs in~\eqref{GnHyp} are multiplied by non-zero
powers of~$\mathfrak{u}$. This means that the time derivatives
contained in~$C_n$ may hit that~$\mathfrak{u}$, thus allowing the log
to survive. Since~$C_n$ is quadratic in time derivatives of the good
field, we can only expect the power of logs to increase. Once again
using computer algebra to compute the first few~$B_n$ and gain some
intuition, we assume that,
\begin{align}\label{BnHyp}
  B_n=\sum_{j=0}^{n-1}\sum_{q=p-n}^{j}\sum_{i=1}^{j+1}A_{j,q,0}(\theta^A)
  \frac{\mathfrak{u}^q}{R^j}\log^i\left(-\frac{R}{\mathfrak{u}}\right)\,,
\end{align}
and try to show that this implies the same statement holds
for~$B_{n+1}$. The first step of this is to check
that~\eqref{BnIDTerms1} and~\eqref{BnIDTerms2} are in the class of
terms in~\eqref{BnHyp} for~$n\rightarrow n+1$, which is
straightforward. Then we plug~\eqref{BnHyp}
into~\eqref{BnsecondInt}. Differentiating and integrating the first
term, and rearranging terms with different angular coefficients we
get,
\begin{align}\label{eq:bad_Mink}
  B_{n+1}=\sum_{j=0}^{n}\sum_{q=p-n-1}^{j}\sum_{i=1}^{j+1}A_{j,q,0}(\theta^A)
  \frac{\mathfrak{u}^q}{R^j}\log^i\left(-\frac{R}{\mathfrak{u}}\right),
\end{align}
as desired. One can trace back that the reason for the logs in the bad
field~\eqref{eq:bad_Mink} to appear as~$log^i(-R/\mathfrak{u})$
with~$i \geq 1$ is that in the logarithmic terms in the good
field~\eqref{eq:bad_Mink} are of the
form~$f(\mathfrak{u},R,\theta^A)\log(-R/\mathfrak{u})$ where~$f$ is
polynomial in~$R$ and~$\mathfrak{u}$.  This differs with the result
reported in~\cite{DuaFenGasHil23} as the logs in the bad field there
only enter linearly and the logs in the good are of the
form~$f(R,\theta^A)\log(-R/\mathfrak{u})$.  However the initial data
used in~\cite{DuaFenGasHil23} for the analysis of the wave equation
with conformal methods is more restricted than that considered here in
the sense that~$M_{\phi,n}$ and~$\bar{M}_{\phi,n}$ contain only a
finite number of spherical harmonics at each order in~$R$. As a sanity
check for the heuristic approach followed in this article, we verified
for a few orders in~$n$ that if we consider the same restriction in
initial data the logarithmic terms in the bad field become linear.
Nonetheless we do not have a proof that is the case for general~$n$.
We can now reorder the terms in~\eqref{BnHyp} by putting together
terms that share powers of~$R$,
\begin{align}
  \label{BReordered}
  &b = \sum_{n=1}^{\infty}\sum_{k=0}^{n} \frac{(\log R)^k 
    \mathcal{B}_{n,k}(\psi^*)}{R^n}\,,
\end{align}
where~$\mathcal{B}_{n,k}(\psi^*)$ are functions that do not vary along
integral curves of~$\psi^a$.

\subsection{Ugly field}

The ugly field satisfies the equation,
\begin{align}
  \mathring{\square} u = \tfrac{2p}{R}\nabla_T u \,,
\end{align}
where~$p$ is an integer. We choose~$p=0$ we turn this equation into a
good. Rescaling~$u$ as,
\begin{align}
  U=uR\,,
\end{align}
we find that the ugly equation turns into,
\begin{align}
  \nabla_\psi\nabla_T (R^pU)=\frac{R^p}{2}(\nabla_\psi^2+\cancel{\Delta})U\,.
\end{align}
We rewrite the field~$U$ as a sum of fields,
\begin{align}
  U=\sum_{n=1}^{\infty}U_n\,.
\end{align}
to get,
\begin{align}\label{ugly-inf-sum}
  \sum_{n=1}^{\infty}\nabla_\psi\nabla_T (R^pU_n)
  =\frac{R^p}{2}\sum_{n=1}^{\infty}(\nabla_\psi^2+\cancel{\Delta})U_n\,.
\end{align}
If we equate terms on either side in the exact same way as in the last
two subsections, we get the following equation,
\begin{align}
  \nabla_\psi\nabla_T (R^pU_n)
  =\frac{R^p}{2}(\nabla_\psi^2+\cancel{\Delta})U_{n-1}\,,
\end{align}
which we can integrate to find an equation for~$U_n$,
\begin{align}\label{UnSecondInt}
	U_n =& \frac{1}{2R^p}\int_{-R}^u\int_{-\mathfrak{u}'}^R
        R'^p(\nabla_\psi^2+\cancel{\Delta})U_{n-1}dR'd\mathfrak{u}'\\ &+
        \int_{-R}^\mathfrak{u}\frac{1}{R^p}\left[\nabla_T (R^p
          U_n)\right]\rvert_{R\rightarrow
          -\mathfrak{u}'}d\mathfrak{u}'+U_n\rvert_{\mathfrak{u}\rightarrow -R}\,.\non
\end{align}
Let us analyze the case~$n=1$ first. This is,
\begin{align}
  U_1 = \int_{-R}^\mathfrak{u}\frac{1}{R^p}\left[\nabla_T (R^p
  U_1)\right]\rvert_{R\rightarrow
  -\mathfrak{u}'}d\mathfrak{u}'+U_1\rvert_{\mathfrak{u}\rightarrow -R}\,.\non
\end{align}
Once more splitting the initial data among the~$U_n$ functions in the
same way as for~$G_n$ and~$B_n$, we find,
\begin{align}\label{U1}
  U_1=\frac{M_{u,1}}{p}\left[1-\frac{(-\mathfrak{u})^p}{R^p}\right]+\bar{M}_{u,1}\,,
\end{align}
for all $p>0$. Notice that there are no logs for $n=1$ in the ugly
equation, but that ceases to be true for higher $n$. In fact, where
the first logs appear depends on the parameter $p$. For general $n$,
the terms coming from initial data become,
\begin{align}
  \label{UnIDTerms1}
  \begin{split}
    &\int_{-R}^\mathfrak{u}\frac{1}{R^p}\left[\nabla_T (R^p
      U_n)\right]\rvert_{R\rightarrow -\mathfrak{u}'}d\mathfrak{u}''=\\ &=
    \begin{cases}
      \frac{M_{u,n}}{p-n+1}\left[\frac{1}{R^{n-1}}
        -\frac{(-\mathfrak{u})^{p-n+1}}{R^p}\right]
      & p-n\neq -1\\ \frac{M_{u,n}}{R^p}\logRu & p-n=-1
    \end{cases}
    \,,
  \end{split}
\end{align}
and,
\begin{align}\label{UnIDTerms2}
	U_n\rvert_{\mathfrak{u}\rightarrow -R} = \frac{\bar{M}_{u,n}}{R^{n-1}}\,.
\end{align}
In order to find the functional form of~$U_n$, we need a hypothesis
similar to~\eqref{BnHyp}. As it turns out, exactly the hypothesis we
used for the bad equation works for the ugly one as well,
\begin{align}\label{UnHyp}
	U_n=&A_{0,0,0}(\theta^A)+\sum_{j=1}^{n-1}
              \sum_{q=j-n}^{j}A_{j,q,0}(\theta^A)\frac{\mathfrak{u}^q}{R^j}\\ 
            &+\sum_{j=1}^{n-1}\sum_{q=0}^{j}\sum_{i=1}^jA_{j,q,i}(\theta^A)
              \frac{\mathfrak{u}^q}{R^j}\log^i
              \left(-\frac{R}{\mathfrak{u}}\right)\,.\non
\end{align}

\subsection{The influence of $p$}

A point must be made here about the influence of the parameter $p$ in
the appearance of logs. From \eqref{UnIDTerms1} we can see that the
first log appears in the terms coming from initial data
when~$n=p+1$. Additionally, we have seen in~\eqref{U1} that logs do
not appear at~$n=1$, so a way to see at which~$n$ the first log
appears is to plug the first term on the right-hand side
of~\eqref{UnHyp} in the first term on the right-hand side
of~\eqref{UnSecondInt} and check under which conditions it generates
logs,
\begin{align}\label{intRp}
  & \frac{1}{2R^p}\int_{-R}^\mathfrak{u}\int_{-\mathfrak{u}'}^R
    R'^p(\nabla_\psi^2+\cancel{\Delta})A_{j,q,0}
    \frac{u'^q}{R'^j}dR'd\mathfrak{u}'
\end{align}
The first integral can only pick up a logarithm if~$p-j=1$. But if~$p$
is large enough that the terms coming from initial data do not
generate logs, namely~$n<p+1$, then this condition is not satisfied
and the first integral does not pick up a log. On the other hand, the
second integral creates a log only if~$p-j+q=-2$ or~$q=-1$. Again, at
any given~$n$, we can make sure the first of these conditions is not
verified by choosing~$p$ sufficiently large. So, if~$p$ is large
enough and~$q\neq -1$, then~\eqref{intRp} becomes,
\begin{align}
  \bar{A}\frac{\mathfrak{u}^{q+1}}{R^{j+1-p}}
  +\tilde{A}\mathfrak{u}^{p+q-j}+A'\frac{1}{R^{j-q}}\,.
\end{align}
Looking at the first term we see that each time a term gets
differentiated and integrated like~\eqref{intRp}, the power
of~$\mathfrak{u}$ in that term increases by one. Since~\eqref{U1} only
has positive powers of~$\mathfrak{u}$, that means~$q$ is never
negative and hence~$q\neq -1$. In other words, the first log is only
generated by terms coming from initial data~\eqref{UnIDTerms1} and
always appears divided by~$R^p$. That means that the higher the
parameter~$p$ in the ugly equation, the higher the power of~$R$ that
divides it. Mathematically we can write this as,
\begin{align}\label{polyhomosuperuglies}
  u =\frac{\mathcal{U}_{1,0}(\theta^A)}{R}+\sum_{n=2}^{p}
  \frac{\mathcal{U}_{n,0}(\psi^*)}{R^n}+\sum_{n=p+1}^{\infty}\sum_{k=0}^{N_n^u}
  \frac{(\log R)^k \mathcal{U}_{n,k}(\psi^*)}{R^n}\,.
\end{align}

\subsection{Good-bad-ugly system}

We are now in a position to put all these results together in one
theorem:
\begin{thm}
  Let~$X^{\ul{\alpha}}=(T,X^{\ul{i}})$ be an asymptotically Cartesian
  coordinate system with an associated covariant derivative~$\mn$ and
  let $\mathcal{S}$ be a Cauchy hypersurface defined by the
  condition~$T=0$. The \textit{good-bad-ugly} system defined as,
  \begin{align}
    &\mathring{\square} g = 0\,, \non\\
    &\mathring{\square} b = (\nabla_T g)^2\,,\non\\
    &\mathring{\square} u = \tfrac{2p}{R}\nabla_T u\,,
  \end{align}
  where~$\mathring{\square}:=\eta^{ab}\mn_a\mn_b$ and~$\eta$ is the
  Minkowski metric, admits formal polyhomogeneous asymptotic solutions
  near null infinity of the type,
  \begin{align}
    &g=\sum_{n=1}^\infty\frac{\mathcal{G}_{n,0}(\psi^*)
      +\mathcal{G}_{n,1}(\psi^*)\log R}{R^n}\,,\non\\
    &b =  \sum_{n=1}^{\infty}\sum_{k=0}^{n} \frac{(\log R)^k 
      \mathcal{B}_{n,k}(\psi^*)}{R^n}\,,\\
    &u =\sum_{n=1}^{p} \frac{\mathcal{U}_{n,0}(\psi^*)}{R^n}
      +\sum_{n=p+1}^{\infty}\sum_{k=0}^{N_n^u}
      \frac{(\log R)^k \mathcal{U}_{n,k}(\psi^*)}{R^n}\,,\non
  \end{align}
  with initial data on~${\mathcal{S}}$ of the type,
  \begin{align}
    \phi\rvert_{\mathcal{S}} = \sum_{n=1}^{\infty}
    \frac{\bar{M}_{\phi,n}}{R^n}\,,\quad
    \nabla_T\phi\rvert_{\mathcal{S}}=
    \sum_{n=1}^{\infty}\frac{M_{\phi,n}}{R^{n+1}}\,,
  \end{align}
  where~$\phi$ can be either~$g$, $b$ or~$u$ and~$\bar{M}_{\phi,n}$
  and~$M_{\phi,n}$ are angular functions. This is valid outside a
  compact ball centered at~$R=0$.
\end{thm}

\section{Asymptotically flat spacetimes}
\label{section:AF_spacetimes}

In this section we want to reach results of the same kind as those in
the first part of this work with two major generalizations. The first
consists in using an asymptotically flat metric in the construction of
the wave operators, instead of the Minkowski metric we used
previously. This will be done by considering the metric functions
defined in the geometric setup to depend analytically on the evolved
fields. Naturally, the end goal of this is to have a theorem that is
applicable to the EFE, in which case the metric functions \textit{are}
the evolved fields, but we will keep the specific form of this
dependence general for now. The second generalization we want to
consider is adding arbitrary sums of \textit{stratified null forms},
or SNFs, to the right-hand sides of the good-bad-ugly system. SNFs are
defined as terms that involve products of up to one derivative of the
evolved fields and fall-off faster than~$R^{-2}$ close to null
infinity. Henceforth~$\mathcal{N}_\phi$, where~$\phi$ is the evolved
field that~$\mathcal{N}_\phi$ is associated to, will denominate an
arbitrary linear combination of stratified null forms. As was seen
in~\cite{DuaFenGasHil21}, the fact that the~$\gamma$ functions are
analytic functions of the evolved fields means that we can Taylor
expand them around~$g=b=u=0$, namely,
\begin{align}\label{taylor}
  \gamma(g,b,u)&=\sum_{i=0}^\infty\sum_{j=0}^\infty\sum_{k=0}^\infty
                 \frac{g^i b^j u^k}{i!j!k!}\left(\frac{\p^{i+j+k}\gamma}
                 {\p g^i\p b^j\p u^k}\right)\Big|_{\mathscr{I}^+}\non\\
               &=\frac{\p \gamma}{\p g}\Big|_{\mathscr{I}^+}g
                 +\frac{\p \gamma}{\p b}\Big|_{\mathscr{I}^+}b
                 +\frac{\p \gamma}{\p u}\Big|_{\mathscr{I}^+}u+...\,,
\end{align}
where the second equality uses the fact
that~$\gamma|_{\mathscr{I}^+}=0$, because the metric is asymptotically
flat. We can also easily show that~\eqref{decayScri} implies that,
\begin{align}
  \gamma=o^+(1)\,.
\end{align}
Another argument made in~\cite{DuaFenGasHil21} that we need here is
that if the evolved fields can be written as polyhomogeneous
functions, then so can the~$\gamma$ functions and the stratified null
forms. We rescale~$g$, $b$ and~$u$ and rewrite them as infinite sums
of functions~$G_n$, $B_n$ and~$U_n$ exactly as was done in the last
section. Then we plug those sums in the generalized good-bad-ugly
system of equations,
\begin{align}
  & \mathring{\square} g = \mathcal{N}_g\,,\non\\
  &\mathring{\square} b = (\nabla_T g)^2 + \mathcal{N}_b\,,\non\\
  &\mathring{\square} u = \tfrac{2}{R}\nabla_T u + \mathcal{N}_u\,.
  \label{gbu2}
\end{align}
Naturally, this includes replacing the metric functions that
constitute the wave operators with their corresponding~$\gamma$'s and
writing those in terms of the evolved fields. It also includes writing
the SNFs, $\mathcal{N}_\phi$, as functions of those fields. We cannot
do that here because we have not specified those dependencies.

\subsection{A new outgoing null vector}

As stated above, performing the necessary integrations in curved
spacetimes involves an extra step. Upon integrating along~$\p_T{}^a$,
we are effectively `running through' integral curves of~$\psi^a$,
dragging~$Q''$ from~$Q'$ to~$Q$ (see
Fig.~\ref{fig:Int_Fig}). Therefore, the point~$Q''$ is determined by
how far up in the integral curve of~$\p_T{}^a$ we are. However, that
particular relation is fixed by the shape of the integral curves
of~$\psi^a$ which, in curved spacetimes, is not trivial. For this
reason we will use another outgoing null vector field~$\tilde{\psi}^a$
along which to do this integration using the eikonal equation. First
we define the coordinate~$\tilde{\mathfrak{u}}$ as the solution to,
\begin{align}\label{eikonal}
  g^{ab}\nabla_a\tilde{\mathfrak{u}}\nabla_b\tilde{\mathfrak{u}}=0\,.
\end{align}
Then we define~$\tilde{\psi}^a$ as,
\begin{align}
  \tilde{\psi}^a = -\nabla^a \tilde{\mathfrak{u}}\,,
\end{align}
where the index is raised with the inverse spacetime
metric~$g^{ab}$. As~\eqref{flatmetricfuncs} is written in terms
of~$\psi^a$, we need a way to write the new vector
field~$\tilde{\psi}^a$ in terms of~$\psi^a$, $\psib^a$
and~$\sg_b{}^a$. In other words, we need to find the behavior of the
coefficients in,
\begin{align}\label{tildepsiComps}
  \tilde{\psi}^a=\lambda\psi^a+\xi\psib^a+\slashed{\delta}^a\,,
\end{align}
where~$\slashed{\delta}^a = \sg_b{}^a\delta^b$. We want to
replace~$\psi^a$ on the right-hand side of~\eqref{flatmetricfuncs}
with~$\tilde{\psi}^a$, $\psib^a$ and~$\sg_b{}^a$. For that we just
need to find out the leading order behavior of the
coefficients~$\lambda$, $\xi$ and~$\delta^a$. Taking~\eqref{eikonal}
and hitting it with a~$\nabla_{\tilde{\psi}}$ derivative, we get,
\begin{align}\label{diffeikonal}
  (\nabla_{\tilde{\psi}}\tilde{\psi})^\alpha=0\,.
\end{align}
Contracting~\eqref{diffeikonal} with~$\sigmab_\alpha$ and taking our
definition of an asymptotically flat metric, a rather long calculation
yields,
\begin{align}
  \nabla_{\tilde{\psi}}\lambda \simeq \frac{1}{2}\nabla_{\psib}\cpr
  +\frac{1}{2}\nabla_{\psi}\cmr-\nabla_\psi\varphi\,.
\end{align}
Integrating this along integral curves of~$\tilde{\psi}^a$ and
choosing the parameter to be~$R$ we get,
\begin{align}\label{lambdaInt}
  \lambda\simeq \int^R_{R_Q}\left(\frac{1}{2}\nabla_{\psib}\cpr
  +\frac{1}{2}\nabla_{\psi}\cmr-\nabla_\psi\varphi\right)dR' 
  + \lambda|_{\mathcal{S}}\,,
\end{align}
where~$R_Q$ depends on the initial data we choose for the derivatives
of~$\tilde{\mathfrak{u}}$. In Minkowski spacetime we had
that~$R_Q=-\mathfrak{u}$ and we can have that now if we require
that~$\tilde{\mathfrak{u}}|_{\mathcal{S}}=-R$. This fixes the spatial
derivatives of~$\tilde{\mathfrak{u}}$ and we can get the time
derivative from the eikonal equation. From~\eqref{tildepsiComps} we
have,
\begin{align}
  \lambda|_{\mathcal{S}}&=\left(-\frac{1}{\tau}
  \tilde{\psi}^{\sigmab}\right)_{\mathcal{S}}\simeq 1\,.
\end{align}
Looking at the integral on the right-hand side we can see that the
first term can generate a log to leading order if we consider that bad
derivatives do not improve the decay of $\cpr$. However, we have
assumed that they do (see Section~\ref{subsection:Assumption}
above). The other two terms are good derivatives and hence can never
contribute to leading order, meaning that we have~$\lambda\simeq
1$. Let us now turn our focus
to~$\delta^a$. Contracting~\eqref{diffeikonal} with~$\sg_\alpha{}^a$
one can compute that,
\begin{align}
  \nabla_{\tilde{\psi}}\slashed{\delta}^a
  \simeq \sg^{aA}\nabla_\psi \mathcal{C}^+_A\,.
\end{align}
Integrating in exactly the same way as we did for~$\lambda$ we find,
\begin{align}
  \slashed{\delta}^a\simeq \int^R_{-\mathfrak{u}}\sg^{aA}\nabla_\psi 
  \mathcal{C}^+_AdR'+\slashed{\delta}^a|_{\mathcal{S}}\,,
\end{align}
and from~\eqref{tildepsiComps} we get,
\begin{align}
  \slashed{\delta}^a|_{\mathcal{S}}\simeq 
  -\sg^{aA}\mathcal{C}_A^+|_{\mathcal{S}}\,.
\end{align}
This means that~$\slashed{\delta}^a=o^+(R^{-1})$. The only remaining
component of~$\tilde{\psi}^a$ is $\xi$, which can be straightforwardly
computed by solving the eikonal equation for~$\xi$. We get,
\begin{align}
  \xi = \frac{e^{-\varphi} \slashed{\delta}^2}{2\tau \lambda} 
  = o^+\left(\frac{1}{R}\right)\,.
\end{align} 
This will allow us to integrate the evolution equations in a similar
way as in the flat spacetime case.

\subsection{Integration}

We can now use the new outgoing null vector field~$\tilde{\psi}^a$ to
modify the expansion of the wave operator~\eqref{lhs2}. The
introduction of~$\tilde{\psi}^a$ is for the sole purpose of making the
integration simpler. Therefore we only need to change the following
term,
\begin{align}
  \nabla_\psi\nabla_T\phi = \lambda^{-1}(\nabla_{\tilde{\psi}}\nabla_T\phi
  -\xi\nabla_{\psib}\nabla_T\phi-\slashed{\delta}^a\nabla_a\nabla_T\phi)\,,
\end{align}
where we have used~\eqref{tildepsiComps}. Note that the second and
third terms on the right-hand side of this equation contain two bad
derivatives, whereas the first one contains one good and one
bad. However, crucially, the coefficients~$\xi$
and~$\slashed{\delta}^a$ decay two orders faster than~$\lambda$, which
means that we can expect the second and third terms to decay one order
faster than the first. We use~\eqref{lhs2} with this modification to
expand the wave operators on the left-hand sides of~\eqref{gbu2}, but
suppress the resulting equations, as they are long and not particularly
enlightening. Now we split our solutions in the exact same way as in
the flat case, namely,
\begin{align}\label{split2}
  \Phi=\sum_{n=1}^{\infty}\Phi_n\,,
\end{align}
and plug that into~\eqref{gbu2}. The next step is to pick a way to
collect terms to equate so that~\eqref{split2} are solutions to the
full equations. As mentioned in the previous section, there are
infinite ways to do that and we choose one that is determined by
Taylor expanding all the~$\gamma$ functions around~$g=b=u=0$ and
organizing all terms by the order in~$R^{-1}$ that they first
contribute to. In the first set of equations we equate those terms
applied to~$\Phi_1$ and we find that to be identical to their flat
counterparts,
\begin{align}\label{G1B1U1eq}
  &\nabla_{\tilde{\psi}}\nabla_TG_1 =0\,,\non\\
  &\nabla_{\tilde{\psi}}\nabla_T B_1=\frac{1}{2R}(\nabla_TG_1)^2 \,,\\
  &\nabla_{\tilde{\psi}}\nabla_T(R^pU_1) = 0\,.\non
\end{align}
In the second set of equations, we equate these terms applied
to~$\Phi_2$ and all the terms that contribute to the next order down
in~$R^{-1}$ applied to~$\Phi_1$, and so on. Doing this systematically
yields the following system of equations,
\begin{align}\label{recur1}
  &\nabla_{\tilde{\psi}}\nabla_TG_n =\Omega^g_{n-1} \,,\non\\
  &\nabla_{\tilde{\psi}}\nabla_TB_n +\frac{1}{2R}C_n=\Omega^b_{n-1} \,,\\
  &\nabla_{\tilde{\psi}}\nabla_T(R^pU_n) = \Omega^u_{n-1}\,,\non
\end{align}
where on the right hand sides~$\Omega^{\phi}_{n-1}$ depend on the
functions~$\{G_{m,k},B_{m,k},U_{m,k}\}$, for~$m\in[1,n-1]$
and~$k\in[0,m]$, and their
derivatives. Also,~$\Omega^{\phi}_{0}:=0$. Note that if all the
equations~\eqref{recur1} are satisfied, then the sums of the
functions~$G_n$, $B_n$ and~$U_n$ are a solution to the good-bad-ugly
system. We begin by integrating~\eqref{recur1} for~$n=1$ in exactly
the same way as in the last section and with the same assumptions on
initial data~\eqref{ID} and~\eqref{IDsplit}. We get,
\begin{align}\label{G1B1U1}
  &G_1 = \bar{M}_{g,1}+M_{g,1}\logRu \,,\non\\
  &B_1=\frac{M_{g,1}^2}{2}\left(\frac{1}{\mathfrak{u}}
    -\frac{1}{R}\right)-\frac{M_{g,1}^2}{2u}\logRu\non\\ 
  &\quad\quad\quad+\bar{M}_{b,1}+M_{b,1}\logRu\,,\non\\	
  &U_1=\frac{M_{u,1}}{p}\left[1-\frac{(-\mathfrak{u})^p}{R^p}\right]
    +\bar{M}_{u,1}\,,
\end{align}
which is exactly the same result as in the flat case. This is to be
expected, since the contributions from the metric curvature and the
SNFs are all higher order and hence do not appear in the~$n=1$
equations. At this point we need to introduce an induction hypothesis
that is in agreement with~\eqref{G1B1U1}. Let us pick the following:
for every~$n$ there is a natural number~$N_n$ such that we can write,
\begin{align}\label{HYP}
  G_n=&\sum_{j=0}^{N_n}\sum_{q=N_n}^{j}
        \sum_{i=0}^{3j+1}A^{(g)}_{j,q,i}(\theta^A)
        \frac{\mathfrak{u}^q}{R^j}\log^i
        \left(-\frac{R}{\mathfrak{u}}\right)\,,\non\\
  B_n=&\sum_{j=0}^{N_n}\sum_{q=N_n}^{j}
        \sum_{i=0}^{3j+1}A^{(b)}_{j,q,i}(\theta^A)
        \frac{\mathfrak{u}^q}{R^j}\log^i
        \left(-\frac{R}{\mathfrak{u}}\right)\,,\\
  U_n=&A^{(u)}_{0,0,0}(\theta^A)+\sum_{j=1}^{N_n}
        \sum_{q=N_n}^{j}\sum_{i=0}^{3j+1}
        A^{(u)}_{j,q,i}(\theta^A)
        \frac{\mathfrak{u}^q}{R^j}\log^i
        \left(-\frac{R}{\mathfrak{u}}\right)\,,\non
\end{align}
where the expression for~$U_n$ is only valid for~$n>1$, since no logs
are allowed in~$U_1$. The most significant departure from the
hypotheses we picked for the flat spacetime case are consequences of
the fact that all equations are now allowed to contain
non-linearities. One such consequence is the fact that now the good
field, in general, has higher powers of logs. Another is that all
three expressions can now contain terms that have logs multiplied by
negative powers of~$\mathfrak{u}$. Additionally, note that from~$n=2$,
all fields may have a power of log of up to four. That is because of
the possibility of there existing terms like~$b^3$ in the SNFs. Before
integrating~\eqref{recur1} we must know the functional form
of~$\Omega^\phi_{n-1}$ that is induced by~\eqref{HYP}. Note that the
most important feature of these expressions is the relationship
between the power of~$R^{-1}$ and~$\log(-R/\mathfrak{u})$.  We can
write the~$\Omega$ functions in the following way,
\begin{align}\label{Omega}
  \Omega^\phi_{n-1} = \sum_{j=2}^{N_n'}\sum_{q=N_n'}^{j}
  \sum_{i=0}^{3j+1}A^{(\phi)}_{j,q,i}(\theta^A)\frac{\mathfrak{u}^q}{R^j}
  \log^i\left(-\frac{R}{\mathfrak{u}}\right)\,,
\end{align}
where~$N_n'$, in general, is a natural number different from~$N_n$. It
may seem awkward that we make no effort to specify these numbers in
our proof. The reason for that is that their only functions are to
bound the power of~$R^{-1}$ above and the power of~$u$ below. The
first is of no importance in our analysis of the behavior of the
evolved fields as we approach null infinity, whereas negative powers
of~$\mathfrak{u}$ only matter to the extent that they are able to
generate logs upon integration. In other words, we are only interested
in the behavior in $u$ to the extent that terms with~$q=-1$ are
possible. Plugging~\eqref{Omega} in~\eqref{recur1} and assuming
that~\eqref{HYP} holds for~$n\rightarrow n-1$ we can finally integrate
our system of equations in exactly the same way as in the case of a
flat spacetime. A straightforward calculation yields a result that can
be written as~\eqref{HYP}, which is what we wanted. Our
hypothesis~\eqref{HYP} holds for arbitrary~$n$. Much in the same way
as in the previous section, recognizing that~$n$ does not have any
clear physical meaning, we can reorganize the terms in~\eqref{HYP} so
that they can be written as a sum over the power of~$R^{-1}$ in the
following way,
\begin{align}\label{gbuReordered}
  &g = \sum_{n=1}^{\infty}\sum_{k=0}^{3n-2} \frac{(\log R)^k 
    \mathcal{G}_{n,k}(\tilde{\psi}^*)}{R^n}\,,\non\\
  &b = \sum_{n=1}^{\infty}\sum_{k=0}^{3n-2} \frac{(\log R)^k 
    \mathcal{B}_{n,k}(\tilde{\psi}^*)}{R^n}\,,\non\\
  &u = \frac{\mathcal{U}_{1,0}(\theta^A)}{R}+ \sum_{n=2}^{\infty}
    \sum_{k=0}^{3n-2} \frac{(\log R)^k 
    \mathcal{U}_{n,k}(\tilde{\psi}^*)}{R^n}\,,
\end{align}
where the functions~$\mathcal{G}_{n,k}(\tilde{\psi}^*)$,
$\mathcal{B}_{n,k}(\tilde{\psi}^*)$
and~$\mathcal{U}_{n,k}(\tilde{\psi}^*)$ do not vary along integral
curves of the new outgoing vector field~$\bar{\psi}^a$.

\subsection{Generalized good-bad-ugly system}

We are now ready to write a generalization of Theorem 1.
\begin{thm}
  Let~$X^{\ul{\alpha}}=(T,X^{\ul{i}})$ be an asymptotically Cartesian
  coordinate system with an associated covariant derivative~$\mn$ and
  let~$\mathcal{S}$ be a Cauchy hypersurface defined by the
  condition~$T=0$. The \textit{good-bad-ugly} system defined as,
  \begin{align}
    & \mathring{\square} g = \mathcal{N}_g\,,\non\\
    &\mathring{\square} b = (\nabla_T g)^2 + \mathcal{N}_b\,,\non\\
    &\mathring{\square} u = \tfrac{2}{R}\nabla_T u + \mathcal{N}_u\,.
  \end{align}
  where~$\mathring{\square}:=g^{ab}\mn_a\mn_b$ and~$g$ is an
  asymptotically flat metric in the sense of~\eqref{decayScri},
  admits formal polyhomogeneous asymptotic solutions near null
  infinity of the type,
  \begin{align}
    &g = \sum_{n=1}^{\infty}\sum_{k=0}^{3n-2} \frac{(\log R)^k 
      \mathcal{G}_{n,k}(\tilde{\psi}^*)}{R^n}\,,\non\\
    &b = \sum_{n=1}^{\infty}\sum_{k=0}^{3n-2} \frac{(\log R)^k 
      \mathcal{B}_{n,k}(\tilde{\psi}^*)}{R^n}\,,\non\\
    &u = \frac{\mathcal{U}_{1,0}(\theta^A)}{R}+ 
      \sum_{n=2}^{\infty}\sum_{k=0}^{3n-2} \frac{(\log R)^k 
      \mathcal{U}_{n,k}(\tilde{\psi}^*)}{R^n}\,,
  \end{align}
  with initial data on~${\mathcal{S}}$ of the type,
  \begin{align}
    \phi\rvert_{\mathcal{S}} = \sum_{n=1}^{\infty}
    \frac{\bar{M}_{\phi,n}}{R^n}\,,\quad
    \nabla_T\phi\rvert_{\mathcal{S}}=
    \sum_{n=1}^{\infty}\frac{M_{\phi,n}}{R^{n+1}}\,,
  \end{align}
  where~$\phi$ can be either~$g$, $b$ or~$u$ and~$\bar{M}_{\phi,n}$
  and~$M_{\phi,n}$ are angular functions. This is valid outside a
  compact ball centered at~$R=0$.
\end{thm}

\subsection{The influence of~$p$}

Recalling that in finding the asymptotic behavior of uglies in flat
spacetimes, the parameter~$p$ determined the order in~$R^{-1}$ at
which the first log would appear, it would be interesting to find out
whether this property is carried over to asymptotically flat
spacetimes. The answer is no, in general. The reason for this has to
do with the fact that our definition for SNFs is quite unspecific and
certain SNFs could potentially generate logs from second order on in
the ugly equations. Fortunately it is not difficult to find out the
properties an SNF must have in order to generate such logs. The
integration of the differential equation that corresponds to~$U_n$
with such an~$n$ that for~$m<n$, $U_m$ contains no logs is,
\begin{align}\label{intRp2}
  & \frac{1}{2R^p}\int_{-R}^\mathfrak{u}\int_{-\mathfrak{u}'}^R
  R'^p(\nabla_\psi^2+\cancel{\Delta})A_{j,q,0}
    \frac{\mathfrak{u}'^q}{R'^j}dR'd\mathfrak{u}'\,.
\end{align}
We saw in the last section that the first integral, namely the one
in~$R$, cannot generate logarithmic terms so long as $p$ is large
enough. We also saw that for the second integral to generate a log, we
would need~$q=-1$. While in the flat spacetime case we could guarantee
that this does not happen below a certain order, we cannot do that
here without specifying the functional form of the SNFs. This,
however, can be done if we restrict our analysis to GR in GHG, for
instance. Upon choosing a specific gauge, we can look at the SNFs and
check whether there exists one that satisfies~$q=-1$. If not, we could
conceivably come up with a result similar to the one we found in flat
spacetimes, which prevents the existence of logs until arbitrarily
high orders in~$R^{-1}$. We leave this for future work, as our main
goal here is to deal with the logs we cannot seem to eliminate except
by choosing very special initial data, the ones that appear in the
good equation.

\section{Reduced Einstein field equations}
\label{section:GHG}

It is possible to write the EFE in GHG as a set of wave equations for
our metric components. This involves a rather long calculation which
has been done in~\cite{DuaFenGas22} and which we will not repeat
here. It suffices to say that each of those equations can be
identified as a good, bad or ugly equation and their character in
these terms depends, to a large extent, on our freedom to choose the
gauge and constraint addition, as arbitrary multiples of the
constraints may be added to the evolution equations before checking a
posteriori that they are satisfied. In order to classify all wave
equations as {\it good}, {\it bad} or {\it ugly}, we need to
rescale~$\mathcal{C}_A^\pm$ to account for the~$R$ factor
in~\eqref{asympflatness}. Therefore we define,
\begin{align}
  R\hat{\mathcal{C}}_A^\pm = \mathcal{C}_A^\pm\,.
\end{align}
We introduce the reduced Ricci tensor
\begin{align}\label{eq:ReducedRicciDefinition}
  \mathcal{R}_{ab} := R_{ab} - \nabla_{(a}Z_{b)}+ W_{ab} \,,
\end{align}
where,
\begin{align}
	Z^a :=\mathbullet{\Gamma}^a + F^a\,,
\end{align}
$\Gb{}^a:=g^{bc}\Gb_b{}^a{}_c$, $F^a$ are the gauge source functions
and~$W_{ab}$ denotes a generic constraint addition where by generic we
mean any expression for which the propagation of the constraints
construction holds. In\cite{DuaFenGas22}, the EFE are derived as
dependent on the reduced Ricci tensor and the gauge freedom and choice
of constraint addition are codified in~$F^a$ and~$Z^a$.

\subsection{Choosing gauge and constraint addition}

Cartesian harmonic gauge is defined as,
\begin{align}\label{Fgeneral}
F^a = \ring{F}^a\,,
\end{align}
where,
\begin{align}\label{eq:GaugeSourcesForCartesianHarmonic}
  \ring{F}^a =
  g^{bc}\Gamma[\mathring{\nabla},\mathbullet{\nabla}]_b{}^{a}{}_c\,.
\end{align}
For clarity, we write down the four components of~$\ring{F}^a$ explicitly,
\begin{align}
  &\ring{F}^\sigma =
  \frac{2e^{-\epsilon}}{R}\cosh h_+\cosh h_\times
  + \mathcal{C}^+_A\ring{F}^A\,,\non\\
  &\ring{F}^{\sigmab} =
  -\frac{2e^{-\epsilon}}{R}\cosh h_+\cosh h_\times
  + \mathcal{C}^-_A\ring{F}^A\,,\non\\
  &\ring{F}^\theta =
  \frac{e^{-\epsilon}\cot \theta}{R^2}e^{h_+}\cosh h_\times
  -\frac{2}{R}\sg^{R\theta}\,,\non\\
  &\ring{F}^\phi =
  -\frac{2e^{-\epsilon}\cot \theta}{\sin \theta R^2}\sinh h_\times
  -\frac{2}{R}\sg^{R\phi}\,.
\end{align}
As each of the components of~$Z^a$ involves a~$\nabla_T$ derivative of
a specific metric function, we can choose the components of~$W_{ab}$
in order to turn~$\mathcal{C}_+^R$, $\epsilon$ and~$\mathcal{C}_A^+$
into ugly fields. One choice of components that does this, and this
choice is highly non-unique, is the following,
\begin{align}\label{ConstAdd}
  &W_{\psi\psi} =
    -\frac{1}{2}(Z^\sigma)^2+\frac{1}{R}Z^\sigma\,,\non\\ 
  &\slashed{W} =
  -\frac{1}{R}Z^{\sigmab}\,,\non\\ 
  &W_{\psi A} = \frac{2}{R}Z^A\,,
\end{align}
with all remaining components set to zero. With these choices, the EFE
in GHG take the following form,
\begin{align}\label{EFECart}
   &\mathring{\square} \varphi = \mathcal{N}_\varphi\,,\non \\
   & \mathring{\square} \mathcal{C}_{+}^R  =
   \frac{2}{R}\nabla_T\mathcal{C}_+^R+
   \mathcal{N}_{\mathcal{C}_{+}^R} \,,\non\\
   &\mathring{\square} \mathcal{C}_{-}^R  =
   - \frac{1}{2}(\nabla_T h_+)^2-\frac{1}{2}(\nabla_T h_\times)^2
   + \mathcal{N}_{\mathcal{C}_{-}^R}\,, \non\\
   &\mathring{\square} \hat{\mathcal{C}}_A^+
   = \frac{2}{R}\nabla_T\hat{\mathcal{C}}_A^+
   +\mathcal{N}_{\mathcal{C}_A^+} \,,\non\\
   &\mathring{\square} \hat{\mathcal{C}}_A^- =
   \frac{4}{R}\nabla_T\hat{\mathcal{C}}_A^-
   +\mathcal{N}_{\mathcal{C}_A^-}\,,\non \\
   &\mathring{\square} \epsilon  =
   \frac{2}{R}\nabla_T\epsilon +\mathcal{N}_{\epsilon}\,,\non\\
   &\mathring{\square} h_+ = \mathcal{N}_{h_+}\,,\non \\
   &\mathring{\square} h_\times = \mathcal{N}_{h_\times}\,.
\end{align}
We can easily see that the fields~$\varphi$, $h_+$ and~$h_\times$
satisfy the good equation, whereas~$\mathcal{C}_-^R$ satisfies the bad
one, $\mathcal{C}_+^R$, $\hat{\mathcal{C}}_A^+$ and~$\epsilon$ satisfy
the ugly equation with~$p=1$ and~$\hat{\mathcal{C}}_A^-$ the ugly
equation with~$p=2$. A minor and straightforward generalization of
Theorem 2, namely considering systems with more than one equation of
each type, allows us to apply it directly to~\eqref{EFECart} and bound
the power of logs at each order in~$R^{-1}$. As seen in the previous
section, goods and bads will have one log contributing to leading
order, while uglies will not.

\subsection{Time symmetric initial data}

In studies of peeling such as~\cite{GasVal18, Val10, Val04a}, time
symmetric initial data is chosen to simplify the calculations. Namely,
initial data on an initial slice~$\mathcal{S}$ such that the extrinsic
curvature~$K_{ij}$ vanishes. It is interesting to find out how this
assumption translates into the language presented here. In the
following we assume time symmetric initial data and, additionally,
that the shift vanishes. Every equation in this subsection is valid
only on~$\mathcal{S}$. We choose to suppress the
subscript~$\rvert_{\mathcal{S}}$ so as to not overload the
notation. We have,
\begin{align}\label{time-symmetry}
  K_{ij}=0 \rightarrow \p_T\gamma_{ij}=0,.
\end{align}
Assuming our class of initial data~\eqref{IDgeneral} in all our metric
functions and using expressions~\eqref{inducedMetric}
and~\eqref{LapseShift}, we can Taylor expand the the resulting
expression from~\eqref{time-symmetry} close to spatial infinity and
keep only the leading contributions. We find the following,
\begin{align}\label{dersGamma}
  &\p_T\gamma_{RR}  \simeq \frac{1}{2}\p_T(2\varphi-\tau)
    \simeq 0\,,\non\\
  &\p_T\gamma_{R\theta^A} \simeq \frac{1}{2}
    \p_T(\mathcal{C}^+_A-\mathcal{C}^-_A)
    \simeq 0\,,\non\\
  &\p_T\gamma_{\theta^1\theta^1} 
    \simeq R^2\p_T(\epsilon+h_+)\simeq 0\,,\non\\
  &\p_T\gamma_{\theta^1\theta^2} \simeq -R^2
    \sin \theta^1\p_Th_\times\simeq 0\,,\non\\
  &\p_T\gamma_{\theta^2\theta^2} 
    \simeq R^2\p_T(\epsilon-h_+)\simeq 0\,,
\end{align}
where in each line the first equality comes from expanding the
components of~$\gamma_{ij}$ and the second comes from requiring that
the extrinsic curvature vanishes on the initial slice. The first and
second equations of~\eqref{dersGamma} give us relationships between
the angular functions~$M_{\phi,1}$, where~$\phi$ is either~$\cpr$,
$\cmr$, $\varphi$, $\mathcal{C}^+_A$ or~$\mathcal{C}^-_A$. The last 3
equations imply that,
\begin{align}\label{Mepshh}
  M_{\epsilon,1}=M_{h_+,1}=M_{h_\times,1}=0\,.
\end{align}
This result, together with~\eqref{G1B1U1} yields that requiring the
extrinsic curvature~$K_{ij}$ to be zero eliminates the leading order
good logs in the metric functions~$\epsilon$, $h_+$ and~$h_\times$ as
we go out to null infinity. We point out that requiring
that~$K_{ij}=0$ and~$\beta_i=0$ are not necessary conditions
for~\eqref{Mepshh} to hold. Those conditions could be relaxed to a
sufficiently fast decay close to spatial infinity.

\section{Peeling}
\label{section:Peeling}

Smooth null infinity implies that peeling is satisfied, which means
that the components of the Weyl tensor decay with certain negative
powers of the radial coordinate as we approach null infinity. In this
section we want to show that our choices of initial data yield an
obstruction to the satisfaction of that property by introducing
fall-offs that involve the logarithm of $R$, which slows down the
otherwise expected decay. The Weyl scalars are 5 complex numbers that
concisely describe the 10 independent Weyl tensor components. They are
obtained from the contraction of the Weyl tensor with unit null
vectors.  Following the conventions of~\cite{Alc08}, we define the
null tetrad (with normalization~$ l_a n^a = - m_a \bar{m}^a = -1 $):
\begin{align}
  &l^a=\frac{\psi^a}{\sqrt{\tau \, e^\varphi}} \,,\non\\
  &n^a=\frac{\ul{\psi}^a}{\sqrt{\tau \, e^\varphi}} \,,
\end{align}
\begin{align}
  &m^a=\frac{1}{\sqrt{2}} \left( e_{(2)}^a + i e_{(3)}^a \right) \,,\non\\
  &\bar{m}^a=\frac{1}{\sqrt{2}} \left( e_{(2)}^a - i e_{(3)}^a \right) \,,
\end{align}
where~$e_{(2)}^a$ and~$e_{(3)}^a$ are real unit vectors orthogonal
to~$l^a+n^a$ and~$l^a-n^a$. The Weyl scalars~$\Psi_4$, $\Psi_3$,
$\Psi_2$, $\Psi_1$ and~$\Psi_0$ are defined as follows:
\begin{align}\label{WeylSExp}
  &\Psi_4=C_{abcd} n^a \bar{m}^b n^c \bar{m}^d \,,\non\\
  &\Psi_3=C_{abcd} l^a n^b \bar{m}^c n^d \,,\non\\
  &\Psi_2=C_{abcd} l^a m^b \bar{m}^c n^d \,,\non\\
  &\Psi_1=C_{abcd} l^a n^b l^c m^d \,,\non\\
  &\Psi_0=C_{abcd} l^a m^b l^c m^d \,,
\end{align}
where~$C_{abcd}$ is the Weyl tensor. We say that the Weyl
scalars~$\Psi_N$ satisfy the peeling property if, for some
appropriately defined radial parameter~$r$, the asymptotic behavior of
the Weyl scalars has the form~\cite{NewPen62}:
\begin{align}\label{Peeling}
  &\Psi_N \sim \frac{1}{R^{5-N}} \,.
\end{align}
In what follows we compute the leading order of $\Psi_4$, $\Psi_3$ and
$\Psi_2$ from the polyhomogeneous asymptotic solutions of the EFE in
GHG. For that we use~\eqref{EFECart} together with Theorem 2.

\subsection{Violation of peeling}

The result we found for the good-bad-ugly system tells us the powers
of logs admitted by asymptotic solutions at each order in~$R^{-1}$ for
general initial data within the class~\eqref{ID}. However, it is
important to notice that while that power is given exactly at leading
order, at subleading orders we only have an upper bound for it. For
instance, our method allows for logs at second order in the good
equation to have a power of up to 4, but knowing the specific
structure of the SNFs and the dependence of the metric upon the
evolved fields could yield a number smaller than that. As it turns
out, leading order logs in some of the equations is sufficient
for~$\Psi_2$ to have an obstruction to peeling. Plugging the
functional form of the asymptotic solutions for the EFE in
GHG~\eqref{EFECart} given by Theorem 2 in the expressions for the Weyl
scalars~\eqref{WeylSExp}, we find that~$\Psi_4$ and~$\Psi_3$ peel,
\begin{align}\label{Weyl}
  &\Psi_4 = \frac{\bar{\Psi}_4(u,\theta^A)}{R}
    +o\left(\frac{1}{R}\right)\,,\non\\
  &\Psi_3 = \frac{\bar{\Psi}_3(u,\theta^A)}{R^2}
    +o\left(\frac{1}{R^2}\right)\,.
\end{align}
We chose to omit the expressions for the
functions~$\bar{\Psi}_n(u,\theta^A)$ because they are extremely long
and not particularly enlightening, and we refer the reader to the
computer algebra notebooks available in~\cite{DuaFenGas25_web}. The
far simpler notation in~\eqref{Weyl} captures the essential features
of those expressions for the purposes of this section. For the
calculation of~$\Psi_2$, we must add a certain quantity of the
constraints. In the usual spirit of numerical evolutions, we do not
assume these to be satisfied a priori, but are free to add any amount
of them to any expression. Let those constraints be defined as the
following contractions of the Ricci tensor,
\begin{align}
  &\mathcal{H} = R_{ab} (l^a + n^a) (l^b + n^b) \,,
\end{align}
\begin{align}
  &\mathcal{M} = R_{ab} (l^a + n^a) (l^b - n^b) \,.
\end{align}
We add the following amount of the Hamiltonian and momentum
constraints, $\mathcal{H}$ and~$\mathcal{M}$, respectively, to the
expression for~$\Psi_2$ and we find,
\begin{align}
  \Psi^\prime_2
  &= \Psi_2 - \frac{\mathcal{H}+\mathcal{M}}{12}  \non\\
  & = \frac{1}{R^3}\sum_{i=1}^{N_i} \bar{\Psi}_{2,i}(u,\theta^A)\log^i R
    +o\left(\frac{1}{R^3}\right)\,,
\end{align}
where~$N_i$ is a natural number that depends on the powers of logs at
first and second order admitted by asymptotic solutions of the EFE, as
discussed above. Because there are goods and bads in~\eqref{EFECart},
then~$N_i$ cannot be smaller than~$1$. This means that there is an
obstruction to peeling in~$\Psi_2$. As shown in~\cite{DuaFenGasHil22a}
and~\cite{DuaFenGasHil23}, a careful choice of gauge and constraint
addition can be used to change the character of some of the EFE in
GHG. Namely, the authors found that there is a gauge in which the 10
EFE turn into a set of 8 uglies and 2 goods ($h_+$ and
$h_\times$). Because uglies do not have logs at leading order, that
would presumably eliminate some of the logs that prevent~$\Psi_2$ from
peeling. However, the leading logs in~$h_+$ and~$h_\times$ contribute
to that obstruction to peeling and there is no gauge that can turn
them into uglies. That means even the choice of gauge explored
in~\cite{DuaFenGas22} and~\cite{DuaFenGasHil22a} yields a leading term
in~$\Psi_2$ that does not peel.

\subsection{Recovering peeling?}

In the last subsection we saw that there is an obstruction to peeling
in~$\Psi_2$ coming from leading order logs in some of the metric
components, and potentially from second order in~$R^{-1}$ in some
others. In this subsection we discuss a potential avenue of further
research into the minimal conditions that would ensure we recover
peeling. We do not aim to prove anything here, but merely point in
what this work seems to suggest is the direction to find such a
result. In order for us to guarantee that the spacetime peels, we must
be able to control 3 types of logarithms. The ones generated by bad
equations, the ones from ugly equations and the ones that appear in
every equation, which we have dubbed 'good logs' simply because of the
fact that they seem to be the only ones that appear in that
equation. Let us address the bad logs first. We have seen
in~\cite{DuaFenGas22} and~\cite{DuaFenGasHil22a} that one can use
gauge freedom to turn the bad equation into an ugly by forcing on the
gauge source functions to satisfy its own wave equation. This allows
for that function to asymptotically cancel out the 'bad part' of the
equation, while keeping hyperbolicity intact. Regarding ugly logs, we
have seen at the end of section IV that the parameter~$p$ in ugly
equations may be enough to push these logs arbitrarily far down in the
expansion close to null infinity. This is surely the case in flat
space with no SNFs, but it is not yet clear if this result carries
over to the EFE in GHG. We leave for future work a careful analysis of
the SNFs in GR to assert whether there exist any in ugly equations
with a single negative power of~$u$ that are capable of generating
undesired logs. Finally, we turn our attention to the good logs. As
far as we know, there is no way to eliminate these logs by means of a
carefully chosen gauge or constraint addition, so we are inclined to
believe that they can only be eliminated through picking special
initial data. The most obvious way in which this would work is by
selecting initial data of compact support. Other possibility is to use
gluing techniques to construct initial data that in a neighborhood of
spatial infinity are exactly Schwarzschild or Kerr initial data.  Such
a choice would definitely not yield good logs at null
infinity. However, it might not be necessary to place so strong an
assumption. In~\cite{GasMagMen24} the authors also obtain
polyhomogeneous solutions for the good field in flat spacetime with no
SNFs.  In that work it is shown that a necessary and sufficient
condition to eliminate the good logs is to impose certain conditions
in the initial data parameters~$M_{\phi,n}$ and~$\bar{M}_{\phi,n}$ as
expressed in terms of spherical harmonics ---see Proposition 1
in~\cite{GasMagMen24}.  Although our method does not give such
detailed control at the level of initial data, the results obtained in
using conformal methods both in the linear and non-linear
setting~\cite{GasMagMen24, GasVal21a, Val03a, BeyDouFra12, Val08,
  Val04} give reason to expect that a regularity condition could be
found with a more refined version of our current asymptotic system
method would allow to identify initial data that does not yield good
logs at null infinity without the very restrictive assumption of data
of compact support or with gluing techniques.  Furthermore even if
one cannot eradicate logs to all orders but simply control the first
order at which they appear one could almost trivially recover
peeling. This is left for future work.

\section{Conclusions}
\label{section:Conclusions}

The principal aim of this work was to show that the asymptotic systems
method can be applied to the EFE in GHG in order to find the positions
of logarithmically divergent terms. That had already been done
in~\cite{DuaFenGas22}, but a crucial logarithm appearing at leading
order in the good equation (standard wave equation) had not yet been
found with this method ---see~\cite{Gas24} for a abridged discussion
on this issue. We began by showing that the good equation in flat
space admits a formal polyhomogeneous asymptotic solution near null
infinity with logs appearing at every order in~$R^{-1}$ linearly. In
order to demonstrate that this log is not a consequence of a very
special choice of initial data, we relaxed our assumptions on the
initial slice and argued that it appears for a much more general class
of data. Then we adapted the method to find asymtotic solutions for
the bad and the ugly equations as well. The bad field shows a leading
order term containing a log whose existence was already
known~\cite{LinRod03,GasGauHil19} and another that is analogous that
found in the good equation. To higher order, due to the non-linearity
in this equation, logs may appear with powers higher than 1. The ugly
field does not allow for any leading logarithm. However, like the bad
field, it may have higher powers of logs further down in the
expansion.

With this knowledge of the Minkowski case we turned our attention to
asymptotically flat spacetimes and generalized the good-bad-ugly
system to include stratified null forms as source terms. The need to
control the limits of integration implied building the outgoing null
vector field along which we integrate from the solution to the eikonal
equation. In flat space, this vector field reduces to that used
earlier. We showed that this generalization of our model admits formal
polyhomogeneous asymptotic solutions near null infinity where the
power of logs in a term proportional to~$R^{-n}$ is bounded above by
$3n-2$. In flat space, the greater the parameter~$p$ in the ugly
equation (see definition in~\eqref{gbu}) the further down in the
expansion the first log appears. In general, this does not carry over
to the case with an asymptotically flat metric and SNFs because there
is a specific type of SNF that can generate such terms regardless
of~$p$. As a corollary to this, given a system with uglies whose SNFs
do not contain such terms, $p$ would indeed push logarithmically
divergent terms down to higher orders.\par We wrote the EFE as wave
equations for a specific set of variables that turns them into a
system of goods, bads and uglies. The knowledge gathered about
asymptotic solutions to the good-bad-ugly model, allowed us to see
that the EFE in GHG imply an obstruction to peeling at~$\Psi_2$. To
the best of our knowledge, this result is in agreement with the
literature ---see for instance~\cite{GasVal18}. We argued furthermore
that as long as the logarithm appearing in the good equations cannot
be eliminated through a clever choice of gauge and constraint
addition, that obstruction can only be removed by carefully
restricting initial data.\par As earlier work shows, the EFE in GHG
can be turned into a system of goods and uglies. It also shows that
there could be requirements on the~$M_{\phi,n}$ angular functions in
the initial data~\eqref{IDgeneral} that will eliminate the logarithm
in the good equation in flat space ---see~\cite{GasMagMen24}. This
suggests that such a condition may exist for the EFE. Additionally, a
closer inspection of the SNFs present in the EFE could potentially
allow us to use the parameter~$p$ to delay the appearance of the first
logs to higher orders in~$R^{-1}$ to such an extent that peeling would
be recovered. This, however, remains to be shown, and we leave it for
future work.

Due to the relatively straightforward generalizability of the
asymptotic systems method to encompass the EFE in GHG, including to
orders beyond the first, we believe the asymptotic systems method to
be a useful tool in finding where the logarithms appear in these
equations. That information will allow us to make better numerical
implementations by eliminating as many logs as possible through
careful gauge choice and constraint addition, thereby only having to
deal with whichever logs remain.

\acknowledgments

MD acknowledges support from the PeX-FCT (Portugal) program
2022.01390.PTDC. JF acknowledges support from the Leung Center for
Cosmology and Particle Astrophysics (LeCosPA), National Taiwan
University (NTU), the R.O.C. (Taiwan) National Science and Technology
Council (NSTC) Grant No. 112-2811-M-002-132, and the European Union
and Czech Ministry of Education, Youth and Sports through the FORTE
project No. CZ.02.01.01/00/22\_008/0004632. EG acknowledges support
from FCT (Portugal) investigator grant 2020.03845.CEECIND and PeX-FCT
2022.01390.PTDC. DH acknowledges support from the PeX-FCT (Portugal)
program 2023.12549.PEX. This work was partially supported by FCT
(Portugal) project No. UIDB/04005/2020.

\normalem
\bibliography{Improved_AsSys_Method.bbl}{}

\end{document}